\documentclass[psfig,usenatbib,useAMS,usegraphicx,subfig]{mn2e}
\usepackage{amssymb}
\usepackage{times}
\usepackage{url}

\begin{document}

\title[Resonance capture]
  {A general model of resonance capture in planetary systems: \\First and second order resonances}

\author[A. J. Mustill \& M. C. Wyatt]
  {Alexander J. Mustill$^1$\thanks{Email: ajm233@ast.cam.ac.uk},
   Mark C. Wyatt$^1$ \\
  $^1$ Institute of Astronomy, University of Cambridge, Madingley Road,
  Cambridge CB3 0HA, UK}

\maketitle

\begin{abstract}
Mean motion resonances are a common feature of both our own Solar System and of extrasolar planetary systems. Bodies can be trapped in resonance when their orbital semi-major axes change, for instance when they migrate through a protoplanetary disc. We use a Hamiltonian model to thoroughly investigate the capture behaviour for first and second order resonances. Using this method,  all resonances of the same order can be described by one equation, with applications to specific resonances by appropriate scaling. We focus on the limit where one body is a massless test particle and the other a massive planet. We quantify how the the probability of capture into a resonance depends on the relative migration rate of the planet and particle, and the particle's eccentricity. Resonant capture fails for high migration rates, and has decreasing probability for higher eccentricities, although for certain migration rates, capture probability peaks at a finite eccentricity. More massive planets can capture particles at higher eccentricities and migration rates. We also calculate libration amplitudes and the offset of the libration centres for captured particles, and the change in eccentricity if capture does not occur. Libration amplitudes are higher for larger initial eccentricity. The model allows for a complete description of a particle's behaviour as it successively encounters several resonances. Data files containing the integration grid output will be available on-line. We discuss implications for several scenarios: (i) Planet migration through gas discs trapping other planets or planetesimals in resonances: We find that, with classical prescriptions for Type~I migration, capture into second order resonances is not possible, and lower mass planets or those further from the star should trap objects in first-order resonances closer to the planet than higher mass planets or those closer to the star. For fast enough migration, a planet can trap no objects into its resonances. We suggest that the present libration amplitude of planets may be a signature of their eccentricities at the epoch of capture, with high libration amplitudes suggesting high eccentricity (e.g., HD~128311). (ii) Planet migration through a debris disc: We find the resulting dynamical structure depends strongly both on migration rate and on planetesimal eccentricity. Translating this to spatial structure, we expect clumpiness to decrease from a significant level at $e\lesssim 0.01$ to non-existent at $e\gtrsim 0.1$. (iii) Dust migration through PR drag: We predict that Mars should have its own resonant ring of particles captured from the zodiacal cloud, and that the capture probability is $\lesssim 25\%$ that of the Earth, consistent with published upper limits for its resonant ring. To summarise, the Hamiltonian model will allow quick interpretation of the resonant properties of extrasolar planets and Kuiper Belt Objects, and will allow synthetic images of debris disc structures to be quickly generated, which will be useful for predicting and interpreting disc images made with ALMA, Darwin/TPF or similar missions.
\end{abstract}


\section{Introduction}

Mean motion resonances (MMRs) occur when two objects' orbital periods are close to a ratio of two integers, and a particular combination of orbital angles, the resonant argument, is librating. Examples in the Solar System include Neptune and Pluto (3:2 resonance) and the inner Galilean moons of Jupiter (4:2:1 Laplace resonance). There are also now numerous examples of suspected or confirmed MMRs in extrasolar planetary systems (e.g., GJ~876~b and c in a 2:1 resonance, \citealt{2001ApJ...551L.109L}).

Mean motion resonances also occur between planets and small dust particles, as seen in the Earth's resonant dust ring~\citep{1994Natur.369..719D}. Some extrasolar debris discs, such as Vega, show evidence of non-axisymmetric clumps \citep{1998Natur.392..788H,2002ApJ...569L.115W}, and several authors have attempted to model these as arising from a planet's resonant perturbations \citep[e.g.,][]{2003ApJ...588.1110K,2003ApJ...598.1321W}.

Although resonant orbits occupy only a small volume of phase space, they are common because of a locking mechanism which can preserve the resonance once attained. If a particle's or planet's orbital semi-major axis changes due to non-conservative forces, the bodies can approach a resonance and then remain trapped there even under further action of the non-conservative forces. The associated orbital angular momentum change then drives an eccentricity change, while the semi-major axis ratio remains approximately fixed at the resonance.

There are many mechanisms by which such a semi-major axis change can be driven. Early work looked at the tidal evolution of satellite orbits \citep{1965MNRAS.130..159G}. In a protoplanetary disc, planets can migrate by tidal interaction with the gas disc \citep[see][for a recent review]{2009AREPS..37..321C}, and small planetesimals by aerodynamic drag \citep{1977MNRAS.180...57W}. In a gas-depleted debris disc, planets can migrate by gravitational scattering of planetesimals \citep{1984Icar...58..109F,2009Icar..199..197K}. Interplanetary dust drifts towards the Sun under the influence of Poynting-Robertson (PR) drag \citep{1979Icar...40....1B}, and large bodies can be moved more slowly by the Yarkovsky effect \citep{2006AREPS..34..157B}. At the end of a star's main-sequence lifetime, planetesimals can experience aerodynamic drag as the star loses mass \citep{2010ApJ...715.1036D}. Moreover, for a planet orbiting the secondary component in a binary system, formation of a disc following mass transfer from the primary to the secondary could trigger renewed planet migration \citep{2010arXiv1001.0581P}.

Resonance capture has been studied by several authors, going back to \cite{1965MNRAS.130..159G}. The regime of adiabatic migration, where the migration timescale is much longer than the resonant argument's libration timescale, has been studied extensively analytically using a Hamiltonian model \citep[e.g.,][]{1982CeMec..27....3H,1984CeMec..32..127B}. With adiabatic migration, capture is certain if the particle has an eccentricity below a critical value, and probabilistic with decreasing probability as eccentricity increases beyond this. Rapid migration was studied using full N-body models by \citet[henceforth W03]{2003ApJ...598.1321W} for the case of a planet migrating into a planetesimal disc, and using the Hamiltonian model by \citet[henceforth Q06]{2006MNRAS.365.1367Q} for general migration scenarios. Q06 obtained capture probability as a function of migration rate and eccentricity for the Hamiltonian containing a single resonant term, and went on to consider the role played by additional resonant terms in affecting capture probability. Such terms can be important when the planet is eccentric.

In this paper we extend this work in a different direction, and using the Hamiltonian model with a single resonant term we calculate capture probabilities, libration amplitudes and offsets for particles that are captured, and eccentricity jumps for those that pass through the resonance without capture.  We validate the model against the numerical integrations of W03 and Q06. We also discuss the application of the model to various migration scenarios, and discuss previous studies in its light, in particular those of \cite{2008A&A...480..551R} who investigated capture of planetesimals by a migrating planet, and \cite{1994Natur.369..719D} who studied the formation of the Earth's resonant dust ring, both using N-body integrations. We should like to emphasize the role eccentricity can play in affecting capture probabilities and libration amplitudes, which, while long understood in the Solar System \citep[e.g.,][]{1999ssd..book.....M}, is sometimes neglected in studies of extrasolar planets and discs.

The Hamiltonian model we use has several advantages over N-body simulations: (1) it allows some results to be derived analytically; (2) it is faster to integrate numerically than the 3-body problem; (3) all resonances of the same order reduce to a Hamiltonian of the same form, with fewer free parameters than the three-body problem. Once a suite of numerical integrations of the Hamiltonian model is performed, it can be applied to any system, without the need for running a different N-body integration every time the system parameters are changed.

The plan of this paper is as follows: In \S\,2 we describe the dimensionless Hamiltonian model. Readers interested in the details may read the Appendix which contains the mathematical derivation. In \S\,3 we summarise how physical parameters relate to the dimensionless parameters for test particles. In \S\,4 we describe the results of our numerical integrations. In \S\,5 we compare the Hamiltonian model to N-body simulations. In \S\,6 we discuss applications. \S\,7 summarises our work.

\section{Description of Hamiltonian model}

For most of this paper we consider the circular restricted three body problem with a massive planet and a massless test particle orbiting a central star with a low mutual inclination. (In the Appendix, we derive suitable formulae for the case of two massive planets orbiting a star, although we do not pursue this further in this paper.) A mean motion resonance occurs when the ratio of two bodies' mean motions is $j:j-k$ where $j$ and $k$ are integers, and when the associated resonant argument $\theta=j\lambda_2-(j-k)\lambda_1-k\varpi$ is librating. Here, $\lambda_i$ are the mean longitudes of the inner and outer bodies, and $\varpi$ is the longitude of pericentre of the test particle\footnote{When we refer to inner and outer bodies, their elements are subscripted $_1$ and $_2$ respectively; when we refer to a planet and a test particle, their elements are subscripted $_\mathrm{pl}$ and unsubscripted respectively.}. The integer $k$ is the order of the resonance; higher order resonances are weaker and often the dynamics are dominated by the low-order resonances. In this paper, we consider resonances of first and second order.

We work with the widely-used Hamiltonian model of mean motion resonances \citep[e.g.,][]{1999ssd..book.....M}. This model is derived by taking the lowest order term in the disturbing function and while suitable for low eccentricities other terms may be important at higher eccentricity \citep[e.g.,][]{2002ApJ...567..596L}. Here we summarise the model and qualitatively describe its behaviour. Readers interested in the mathematical derivation are referred to the Appendix. In \S\,3 we summarise the mathematical results.

\subsection{Absence of migration}

First we consider the case with no migration forces. A particle is in resonance if its resonant argument is librating, typically about $\pi$, rather than circulating. The evolution of the canonical angle $k(\theta+\pi)$ is governed by the Hamiltonian
\begin{equation}
\mathcal{H}=J^2+\beta J + (-1)^kJ^{k/2}\cos k\theta,
\end{equation}
where $J$, the generalised momentum conjugate to $\theta$, is proportional to the square of the particle's eccentricity. There is one parameter, $\beta$, which measures how far the particle is from the nominal resonance location. All resonances of order $k$, of any value of $j$, and whether the particle is interior to or exterior to the planet, can be reduced to this form; the relation between the dimensionless variables and parameters and physical parameters changes, however, as described in the next Section. For first order resonances this is valid for coplanar systems and for low inclinations: there are no first-order inclination terms in the disturbing function. For second order resonances in a non-coplanar system the inclination terms may be important, and capture into a second-order inclination resonance is possible \citep{1999ssd..book.....M}. Capture into such resonances can be handled using the same formalism but with inclination taking the place of eccentricity, and changes to the scaling coefficients between the Hamiltonian model and the physical variables.

\begin{figure}
\includegraphics[width=.5\textwidth]{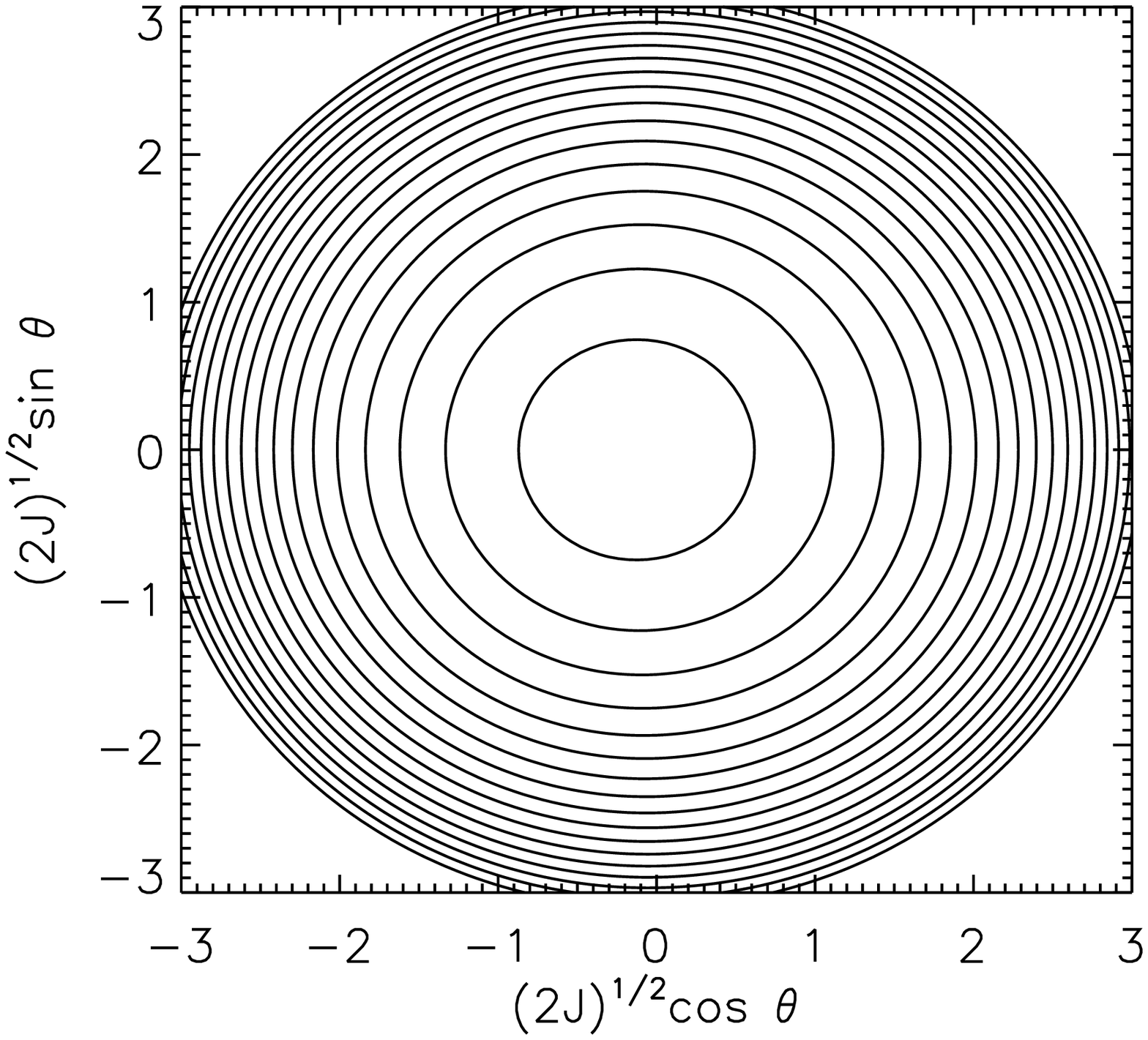}
\includegraphics[width=.5\textwidth]{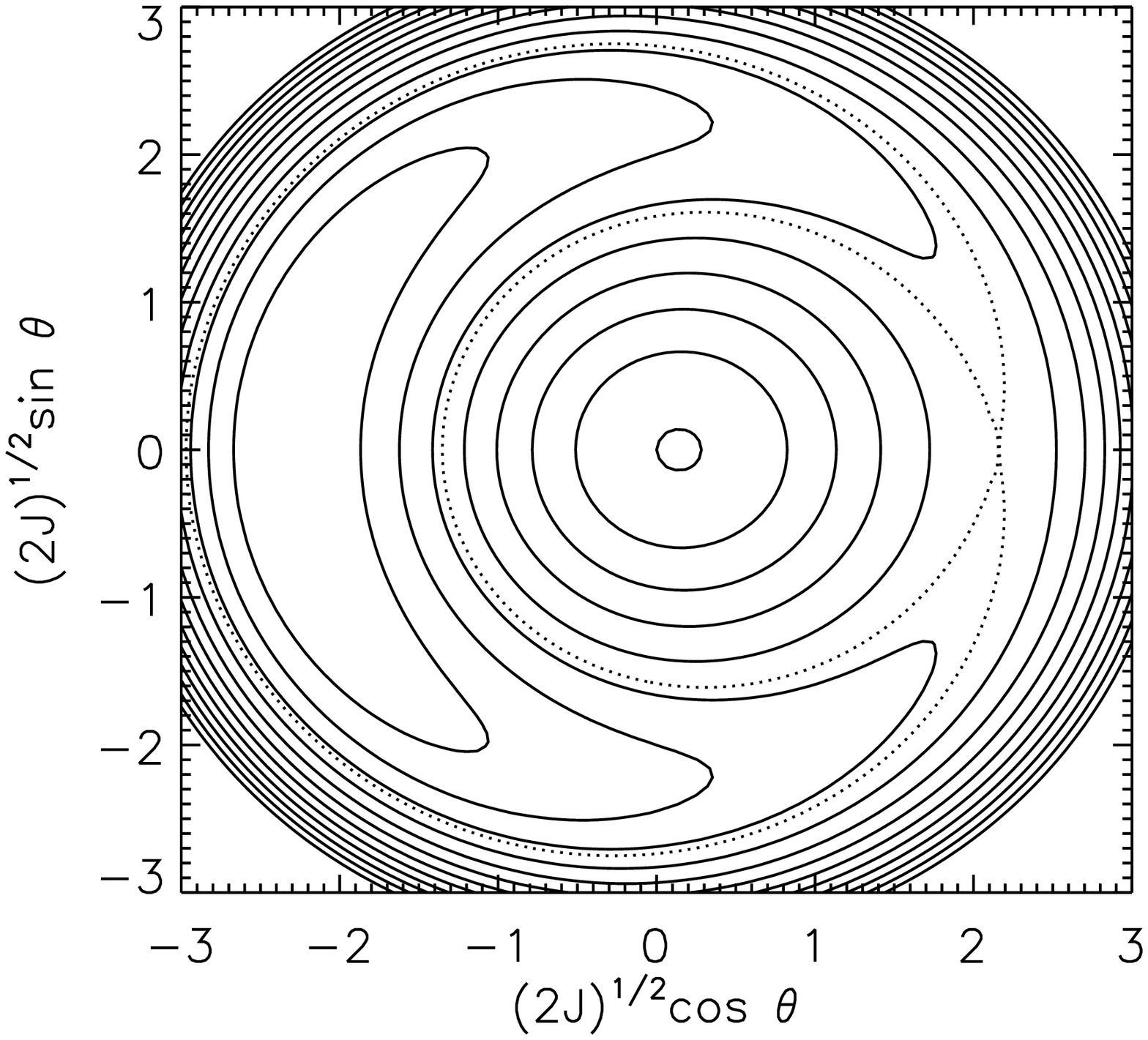}
\caption{Level curves of the Hamiltonian for first-order resonances. In the absence of migration, trajectories are constrained to lie on these curves. The resonant argument is measured clockwise from the origin, and the radial distance is proportional to the eccentricity. \textbf{Top:} prior to resonance crossing, with $\beta$ positive. All trajectories circulate about the origin. \textbf{Bottom:} after resonance crossing, with $\beta$ negative. As well as trajectories where the resonant argument circulates about the origin, librational solutions are possible; these trajectories have been captured into the resonance. The curve demarcating these two regimes is the separatrix (dotted line). The trajectories with small $J$ in the upper panel are now those in the libration region.}
\label{fig:level curves}
\end{figure}

Level curves of the Hamiltonian for $k=1$ (i.e., first-order resonances) are shown in Figure~\ref{fig:level curves}. In this plot, the resonant argument is measured anticlockwise from the positive $x$-axis, and the radius is proportional to $\sqrt{J}$, i.e., to the particle's eccentricity. If $\beta$ is constant, trajectories are confined to these level curves. For $\beta>-3$, there is one fixed point close to the origin, and trajectories circulate about this, so no resonant trajectories exist. For $\beta<-3$, there are stable fixed points both close to the origin, and removed from the origin at $\theta=\pi$. The trajectories librating about this latter point are the resonant trajectories. They may have a larger or smaller libration amplitude. There are also circulating trajectories at smaller and larger eccentricity than the libration region. The curve demarcating these regions is the separatrix, shown as a dotted line in Figure~\ref{fig:level curves}. 

\subsection{Presence of migration}

Now we describe what can happen if there is a non-conservative force acting to change the semi-major axis ratio. Migration of either the inner or outer body, or both, is mathematically identical: only the relative migration rate $\dot{a}_1/a_1-\dot{a}_2/a_2$ matters. Note that capture is only possible for converging orbits \citep{1999ssd..book.....M} because of the geometry of orbital conjunctions \citep{2002ApJ...564L.105C}, so we restrict ourselves to this case.

Prior to reaching the commensurability, $\beta$ is positive, and all trajectories are circulating as shown in the top panel of Figure~\ref{fig:level curves}. Migration reduces the value of $\beta$, until at $\beta=-3$ the separatrix forms and libration in the resonance becomes possible. As $\beta$ continues to decrease, the trajectory may either become trapped in the resonant region, or continue circulating about the origin.

In the case of adiabatic migration, progress can be made analytically by exploiting the invariance of the area enclosed by the trajectory, and the criterion for capture is well understood: if the initial eccentricity is small, the separatrix forms outside the trajectory, and the trajectory is then trapped inside the libration region. Note that the trajectories with small $J$ in the top panel of figure~\ref{fig:level curves} deform continuously into those in the libration region. If the initial eccentricity is large, the separatrix forms inside the trajectory, and initially the angle still circulates. However, continued migration causes the area enclosed by the separatrix to expand, while that enclosed by the trajectory remains constant, so that soon the separatrix hits the trajectory. Then the area enclosed by the trajectory changes and the trajectory jumps to either the libration or the inner circulation region \citep{1999ssd..book.....M}. Although the equations are deterministic, when one has no information on the phase at which the separatrix meets the trajectory, capture can be treated as a probabilistic event, with the probability of capture decreasing as the particle's initial eccentricity is increased \citep{1984CeMec..32..127B}. If the particle is captured into resonance, it remains so under further migration. If the particle is not captured into resonance, circulation continues but with a different eccentricity (which is always smaller for adiabatic migration). If migration continues, however, the particle may be captured into another resonance.

Following capture, the libration continues, with the particle's eccentricity increasing as the migration moves the fixed point further from the origin. The centre of libration is also offset slightly from the case with no migration. Finally, as we show in the Appendix, the libration amplitude decreases slightly, as the angular width of the libration region changes while the area remains fixed. If the migration force is removed, there will no longer be an offset of the libration centre, and the libration amplitude will no longer vary.

The case of rapid migration has been less well explored. Q06 used this Hamiltonian model to derive capture probabilities as a function of both migration rate and initial eccentricity. Here we extend these investigations to more thoroughly map out the parameter space, to find the libration amplitude and offset, and to find eccentricity changes if capture does not occur. This allows us to build up a full picture of what happens as a result of an encounter with a resonance.

Second-order resonances show qualitatively similar behaviour to the first-order resonances. 

\section{Scalings for test particles}

Here we summarise how the dimensionless variables of the Hamiltonian model relate to the physical variables and parameters of a real star--planet--test particle system. The derivation of these equations is given in the Appendix.

\subsection{First-order resonances}

The dimensionless Hamiltonian is
\begin{equation}\label{eq:H simple 1st}
\mathcal{H}=J^2 + \beta J -J^{1/2}\cos \theta.
\end{equation}

For first-order resonances, the dimensionless quantities $J$ and $\dot\beta$, and the dimensionless time $t^\prime$, relate to the physical parameters as follows:
\begin{eqnarray}\label{eq:scalings first}
  J  &=&  k_j\left(\frac{m_\mathrm{pl}}{\mathrm{M}_\oplus}\right)^{-2/3}\left(\frac{m_\star}{\mathrm{M}_\odot}\right)^{2/3}e^2,\\
  \dot\beta &=& -l_j
  \left(\frac{m_\mathrm{pl}}{\mathrm{M}_\oplus}\right)^{-4/3}
  \left(\frac{m_\star}{\mathrm{M}_\odot}\right)^{5/6}\nonumber\\
  &&\left(\frac{a_\mathrm{pl}}{1\mathrm{\,AU}}\right)^{1/2}
  \left[\alpha\frac{\dot{a}_2}{1\mathrm{\,AU/Myr}}-\frac{\dot{a}_1}{1\mathrm{\,AU/Myr}}\right]\\
  t^\prime&=&g_j
  \left(\frac{m_\mathrm{pl}}{\mathrm{M}_\oplus}\right)^{2/3}
  \left(\frac{m_\star}{\mathrm{M}_\odot}\right)^{-1/6}
  \left(\frac{a_\mathrm{pl}}{1\mathrm{\,AU}}\right)^{-3/2}
  \frac{t}{1\mathrm{\,Myr}}.
\end{eqnarray}
Here, $\alpha$ is the ratio of the planet's and particle's semi-major axes: $\alpha=a/a_\mathrm{pl}$ for an internal particle and $\alpha=a_\mathrm{pl}/a$ for an external particle. The coefficients $k_j$ and $l_j$ depend on the particular resonance and are tabulated in Table~(\ref{tab:l_j}). For test particles outside the planet's orbit the ``e'' columns are to be used; for particles inside the planet's orbit the ``i'' columns are to be used.

When capture occurs and migration continues, the eccentricity, libration amplitude and offset evolve with time in the following manner:
\begin{eqnarray}\label{eq:time evolution first}
  e&=&\sqrt{\frac{g_jl_j}{2k_j}}
    \left(\frac{a_\mathrm{pl}}{1\mathrm{\,AU}}\right)^{-1/2}
    \left(\frac{t}{1\mathrm{\,Myr}}\right)^{1/2}\nonumber\\
    &&\left[\alpha\frac{\dot{a}_2}{1\mathrm{\,AU/Myr}}-\frac{\dot{a}_1}{1\mathrm{\,AU/Myr}}\right]^{1/2}\label{eq:e first}\\
  \Delta\theta&\propto&t^{-1/8}\label{eq:amp-t 1st}\\
  \theta_\mathrm{eq}&=&\sqrt{\frac{l_j}{2g_j}}
  \left(\frac{m_\mathrm{pl}}{\mathrm{M}_\oplus}\right)^{-1}
  \left(\frac{m_\star}{\mathrm{M}_\odot}\right)^{1/2}
  \left(\frac{a_\mathrm{pl}}{1\mathrm{\,AU}}\right)\nonumber\\
  &&\left[\alpha\frac{\dot{a}_2}{1\mathrm{\,AU/Myr}}-\frac{\dot{a}_1}{1\mathrm{\,AU/Myr}}\right]^{1/2}
  \left(\frac{t}{1\mathrm{\,Myr}}\right)^{-1/2}
\end{eqnarray}
Eccentricity is pumped up, while the libration amplitude and offset are reduced. Note that the value of the libration amplitude is not determined analytically, but its dependence on time can be.

\begin{table}
  \begin{center}
    \begin{tabular}{ccccccc}
          $j$ & $g_j$ (e) & $k_j$ (e) & $l_j$ (e) & $g_j$ (i) & $k_j$ (i) & $l_j$ (i) \\\hline\\
          2   & 850.788   & 11077.7   & 13.0205   & 3114.07   & 3026.51  & 3.08553\\
          3   & 4797.66   & 5893.36   & 0.818921  & 6003.93   & 4709.31  & 1.02782\\
          4   & 7874.86   & 7180.91   & 0.455939  & 9233.21   & 6124.49  & 0.535696\\
          5   & 11273.1   & 8360.39   & 0.296648  & 12754.1   & 7389.62  & 0.336163\\
          6   & 14942.3   & 9461.19   & 0.211061  & 16527.8   & 8553.59  & 0.233766\\
          7   & 18847.9   & 10500.9   & 0.159184  & 20525.4   & 9642.71  & 0.173547\\
          8   & 22964.5   & 11491.4   & 0.125099  & 24724.9   & 10673.2  & 0.134820\\
          9   & 27272.8   & 12440.7   & 0.101368  & 29108.7   & 11656.0  & 0.108285\\
          10  & 31757.1   & 13355.0   & 0.0841070 & 33662.7   & 12598.9  & 0.0892231\\
          11  & 36404.7   & 14238.9   & 0.0711143 & 38375.3   & 13507.7  & 0.0750161
    \end{tabular}
    \caption{Numerical coefficients for conversion between the physical and dimensionless variables for first order internal (i) and external (e) test particles. Note that the authors believe that any errata in Quillen (2006) have been corrected in this paper.}
    \label{tab:l_j}
  \end{center}
\end{table}

\subsection{Second-order resonances}

For second-order resonances, we write the dimensionless Hamiltonian
\begin{equation}\label{eq:H simple 2nd}
\mathcal{H}=J^2 + \beta J +J\cos 2\theta.
\end{equation}

The scalings for $J$, $\dot\beta$ and $t^\prime$ are then as follows:
\begin{eqnarray}\label{eq:scalings second}
  J  &=&  k_j\left(\frac{m_\mathrm{pl}}{\mathrm{M}_\oplus}\right)^{-1}\left(\frac{m_\star}{\mathrm{M}_\odot}\right)e^2,\\
  \dot\beta &=& -l_j
  \left(\frac{m_\mathrm{pl}}{\mathrm{M}_\oplus}\right)^{-2}
  \left(\frac{m_\star}{\mathrm{M}_\odot}\right)^{3/2}\nonumber\\
  &&\left(\frac{a_\mathrm{pl}}{1\mathrm{\,AU}}\right)^{1/2}
  \left[\alpha\frac{\dot{a}_2}{1\mathrm{\,AU/Myr}}-\frac{\dot{a}_1}{1\mathrm{\,AU/Myr}}\right]\\
  t^\prime&=&g_j\frac{m_\mathrm{pl}}{\mathrm{M}_\oplus}
  \left(\frac{m_\star}{\mathrm{M}_\odot}\right)^{-1/2}
  \left(\frac{a_\mathrm{pl}}{1\mathrm{\,AU}}\right)^{-3/2}
  \frac{t}{1\mathrm{\,Myr}}.
\end{eqnarray}

\begin{table}
  \begin{center}
    \begin{tabular}{ccccccc}
          $j$ & $g_j$ (e)   & $k_j$ (e) & $l_j$ (e) & $g_j$ (i) & $k_j$ (i) & $l_j$ (i) \\\hline\\
          3   & 4.57110     & 557561    & 225528    & 32.5896  & 108449   & 27687.7   \\
          5   & 128.778     & 81348.5   & 852.466   & 146.484  & 120638   & 1543.58   \\
          7   & 305.071     & 75601.1   & 253.169   & 332.276  & 124094   & 373.904   \\
          9   & 553.801     & 72865.1   & 107.555   & 590.374  & 125717   & 143.877   \\
          11  & 874.950     & 71261.2   & 55.4009   & 920.840  & 126658   & 69.8820   \\
          13  & 1268.52     & 70206.1   & 32.2138   & 1323.69  & 127271   & 39.0819   \\
          15  & 1734.49     & 69459.2   & 20.3629   & 1798.92  & 127704   & 24.0293   \\
          17  & 2272.84     & 68903.7   & 13.6835   & 2346.56  & 128024   & 15.8149   \\
          19  & 2883.61     & 68472.8   & 9.63420   & 2966.61  & 128269   & 10.9566   \\
          21  & 3566.83     & 68128.8   & 7.03771   & 3659.06  & 128465   & 7.90129
    \end{tabular}
    \caption{Numerical coefficients for conversion between the physical and dimensionless variables for second order internal (i) and external (e) test particles.}
    \label{tab:l_j 2nd order}
  \end{center}
\end{table}

The coefficients are tabulated in Table~(\ref{tab:l_j 2nd order}).

When capture occurs and migration continues, the eccentricity evolves according to Equation~\ref{eq:e first}; libration amplitude and offset evolve with time in the following manner:
\begin{eqnarray}
  \Delta\theta&\propto&t^{-1/4}\label{eq:amp-t 2nd}\\
  \theta_\mathrm{eq}&=&\frac{1}{4}g_j^{-1}
  \left(\frac{m_\mathrm{pl}}{\mathrm{M}_\oplus}\right)^{-2/3}
  \left(\frac{m_\star}{\mathrm{M}_\odot}\right)^{1/6}
  \left(\frac{a_\mathrm{pl}}{1\mathrm{\,AU}}\right)^{3/2}\nonumber\\
  &&\left(\frac{t}{1\mathrm{\,Myr}}\right)^{-1}
\end{eqnarray}
This is similar to first-order resonances, but the time dependence of the libration amplitude and offset is stronger. Again, the value of the libration amplitude is not determined analytically, but its dependence on time can be.

\section{Numerical integration}

\label{S:numerics}

\subsection{Capture probabilities}

\subsubsection{First order resonances}

\label{capture first order}

For our numerical investigations, we show plots of capture probability, libration amplitude and offset, and eccentricity jump, in a region $\log|\dot{\beta}|\in[-3,1)$, $\log J_0\in[-2,2)$. Here $J_0$ is the initial value of the momentum $J$. Note that higher-order terms in the disturbing function will become important for $e\gtrsim 0.1$, i.e., $J_0\gtrsim 1$ or $100$ for a Jupiter-mass or an Earth-mass planet respectively. Note also that an eccentricity of $1$ is attained at $J_0\sim 100$ or $10^4$, for Jupiter-mass and Earth-mass planets respectively. Hence, results on the extreme right-hand side of the following figures may not be accurate for massive planets.

We consider the Hamiltonian given by Equation~(\ref{eq:H simple 1st}). For the integration we transformed to Poincar\'e's canonical Cartesian variables (equation~\ref{eq:poincare cartesian}) in order to avoid a singularity in the equations of motion for small $J$. The Hamiltonian is then
\begin{equation}\label{eq: H poincare cartesian}
\mathcal{H}=\frac{\left(x^2+y^2\right)^2}{4} + \frac{\beta}{2}\left(x^2+y^2\right) - \frac{x}{\sqrt{2}}.
\end{equation}
The equations of motion arising from this Hamiltonian ($\dot{x}=-\partial\mathcal{H}/\partial y$, $\dot{y}=\partial\mathcal{H}/\partial x$) were integrated numerically with a Runge-Kutta routine with Cash-Karp spacings and adaptive step size \citep{2007nrca.book.....P}. We also vary $\beta$ by including the equation $\dot\beta=\mathrm{constant}$ to impose migration; this changes semi-major axis without affecting eccentricity. Note that it is sufficient to consider linear variation in $\beta$ since it is only the migration rate at the instant of resonant passage that affects capture probability, so long as the migration is sufficiently smooth. However, if the migration rate changes rapidly, such as if the planet is experiencing turbulent torques, then the model does not apply. The initial value of $\beta$ was chosen to be $15\pm 1$ with a uniform random distribution. It was found that when all particles were started from exactly the same value of $\beta$ the capture probability varied in a very rapid manner with $J_0$ or $\dot\beta$. This appears to be related to the phase at which particles enter the resonance; the distribution of phases does not remain uniform as the system evolves away from the initial uniform distribution. While perhaps mathematically interesting, this is not physically useful, so we randomised over this parameter.

\begin{figure}
\includegraphics[width=0.5\textwidth]{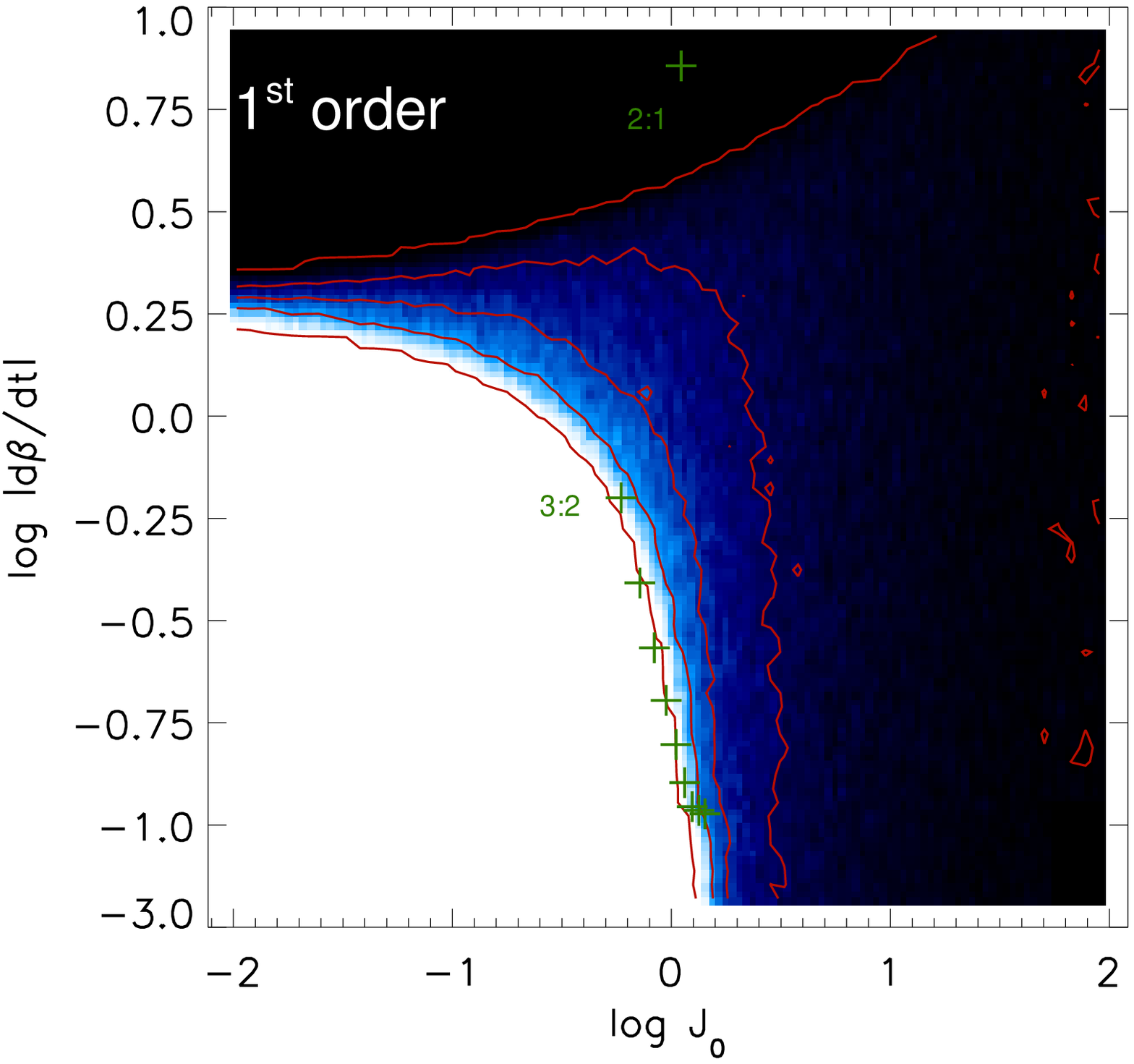}
\includegraphics[width=0.5\textwidth]{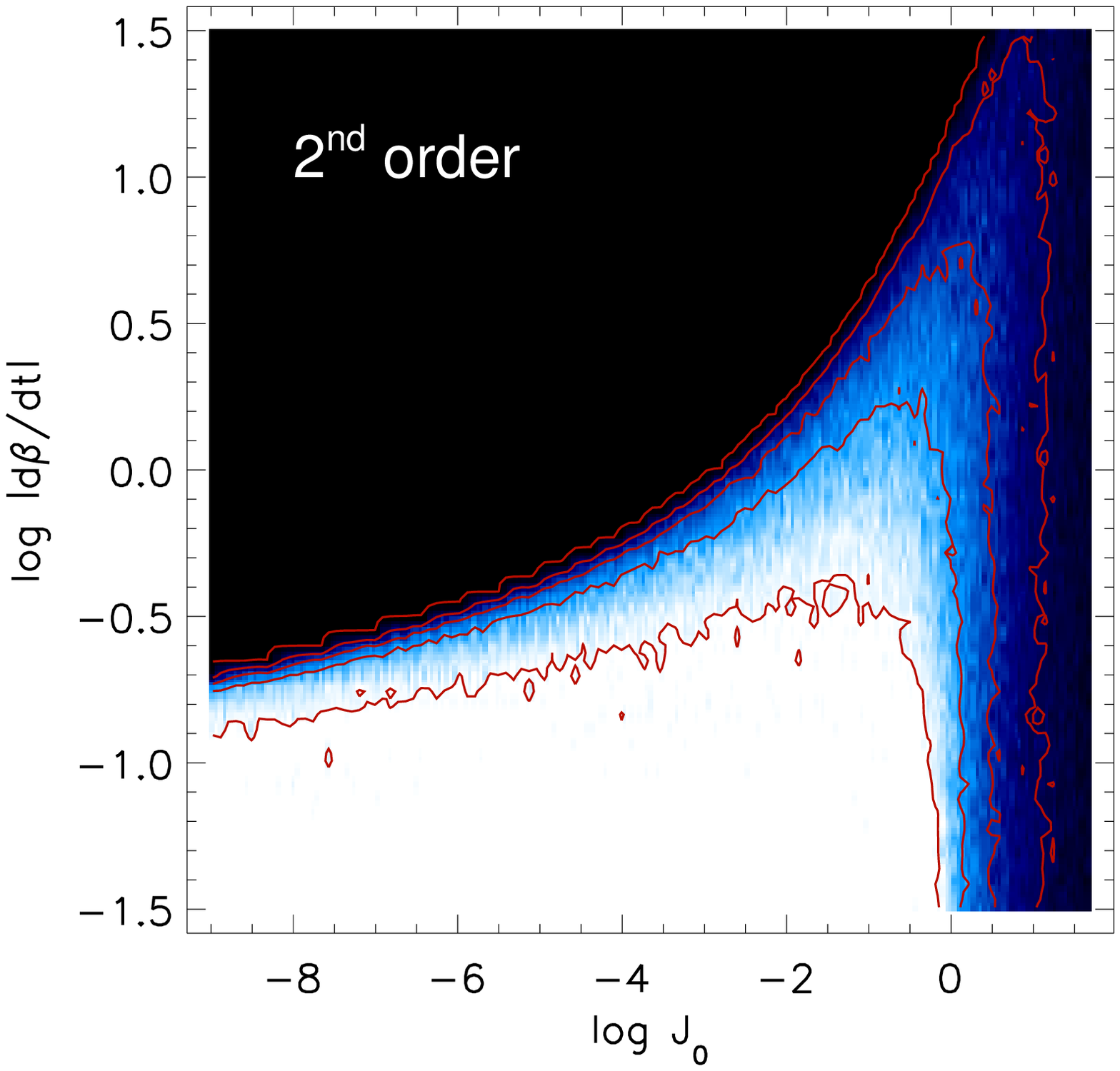}
\caption{\textbf{Top:} Capture probabilities for the first-order resonance, as a function of rescaled eccentricity ($x$-axis) and migration rate ($y$-axis). Contours are at 1\%, 25\%, 50\%, 75\% and 99\% probability. Also shown (green crosses) are the dimensionless migration rate and eccentricity for a particle with Keplerian eccentricity 0.01, migrating at 1\,AU\,Myr$^{-1}$ into exterior resonances with an Earth-mass planet orbiting a Solar-mass star at 1\,AU. The 2:1 and 3:2 resonances are labelled. Note that the axis of ordinates is condensed at the bottom of the plot. The bottom right corner ($\log J_0\in[1.75,2),\log\dot\beta\in[-3,-1)$) was not integrated due to long integration times. 
\textbf{Bottom:} Capture probabilities for second-order resonances. Contours are at 1\%, 25\%, 50\%, 75\% and 99\% probability.}
\label{fig:prob}
\end{figure}

We varied the initial momentum $J_0$ and the migration rate $\dot\beta$. For each point in parameter space we integrated 100 trajectories, with the resonant argument chosen from a uniform distribution over $[0,2\pi)$ and initial $\beta$ as described above. A trajectory was classed as a capture if all of 1000 output values of $\theta$ were within $4\pi/5$ radians of the centre of libration. This avoids any misclassification: choosing a larger libration amplitude would risk misclassifying circulating trajectories as captures. The top panel of Figure~(\ref{fig:prob}) shows capture probabilities as a function of $J_0$ and $\dot\beta$\footnote{The data used to create figure~\ref{fig:prob} and the other other contour plots in this paper are available on-line at \url{http://www.ast.cam.ac.uk/~ajm233/} and the journal website and may be used provided that this work is cited.}.

In Figure~\ref{fig:prob}, top panel, we see that capture into resonance is guaranteed for small initial eccentricities and migration rates. For low migration rates we are in the well-studied adiabatic regime. For low eccentricities capture is certain since the separatrix forms around the initial trajectory. For high eccentricities the separatrix forms inside the initial orbit and expands to meet it as the migration continues. Capture then is probabilistic, with a probability that decreases as the initial eccentricity increases \citep{1982CeMec..27....3H}.

For low eccentricities, with $J_0<1.3$, capture is certain if the migration rate is low and impossible if it is high. The width of the transition region from capture to no capture increases with eccentricity. In the limit of low eccentricity, the transition occurs at a critical migration rate of $|\dot\beta|\approx 2.1$. Certain capture occurs for low migration rates up to $J_0\approx 1.3$. For higher eccentricities ($J_0>1.3$), capture is always probabilistic, with a capture probability that is not strongly dependent of migration rate; however, if migration rate is too high, then capture is still impossible. Interestingly, the maximum migration rate allowing capture is slightly higher for higher eccentricities. This is because this maximum migration rate is governed by the resonant libration period, which decreases with higher $J$ (Eq.~\ref{eq:period 1st}). For a given migration rate above the critical rate, the capture probability peaks at a finite eccentricity.

Vertical line cuts through Figure~\ref{fig:prob}, bottom panel, give the same results as shown in figure~2 of Q06, who integrated the same Hamiltonian with a different integrator, providing a check on our numerical implementation.

In subsequent sections we shall sometimes be concerned with particles that have dimensionless momenta $J_0>100$, for example when considering planets of around Earth mass or lower. To calculate the capture probabilities at these high momenta, we extrapolated the results plotted in Figure~\ref{fig:prob}. At high momentum ($J_0\gtrsim 10$) the capture probability is independent of migration rate for most of the range of migration rates considered, and a regression fit after pooling the data for $J_0\in[3.2,100)$ and $\dot\beta\in[0.001,4.2]$ gives $p=0.499J_0^{-0.725}$ for the capture probability. This agrees with the analytical results for the adiabatic case where $p\propto e^{-3/2}$ \citep{1988Icar...76..295D}.

\subsubsection{Second-order resonances}

We used Poincar\'e variables and same integrator as for the first-order resonance to integrate the second-order Hamiltonian
\begin{equation}
\mathcal{H}=\frac{\left(x^2+y^2\right)^2}{4}+\frac{(\beta+1)x^2}{2}+\frac{(\beta-1)y^2}{2}.
\end{equation}
 Again we randomised the initial angle and distance to the resonance. The integrations were performed for $\log|\dot{\beta}|\in[-1.5,1.5)$, $\log J_0=\in[-9,1.6875)$. Higher order terms will become important for J of several hundred. The capture probabilities are shown in Figure~(\ref{fig:prob}), bottom panel.

The overall picture is similar to that for first order resonances: Certain capture at low migration rate and eccentricity, no capture at high migration rate and low eccentricity, and probabilistic capture at low migration rate and high eccentricity. In the adiabatic limit, the transition from certain to probabilistic capture occurs at $J_0\approx 0.8$. At high eccentricities, capture probability does not depend strongly on migration rate for the rates considered, and we find $p=0.867J_0^{-0.48}$, agreeing with analytical studies \citep{1988Icar...76..295D} which give $p\propto e^{-1}$. In the limit of low eccentricity, the transition from certain to impossible capture does not appear to converge to any value of $\dot\beta$, unlike for first order resonances, being quite strongly dependent on the migration rate, particularly for $J_0\gtrsim 0.001$. At higher eccentricities, the maximum migration rate allowing capture increases more rapidly with $J$ than for first order resonances. This is due to the stronger dependence of the libration period on $J$ (Eq.~\ref{eq:period 2nd}).

Vertical line cuts through Figure~\ref{fig:prob}, bottom panel, give the same results as shown in figure~3 of Q06.

\subsection{Libration amplitude and evolution in resonance}

\begin{figure}
\includegraphics[width=.5\textwidth]{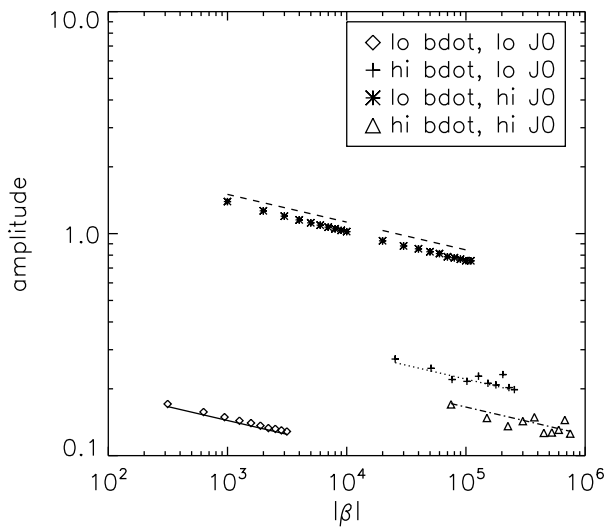}
\includegraphics[width=.5\textwidth]{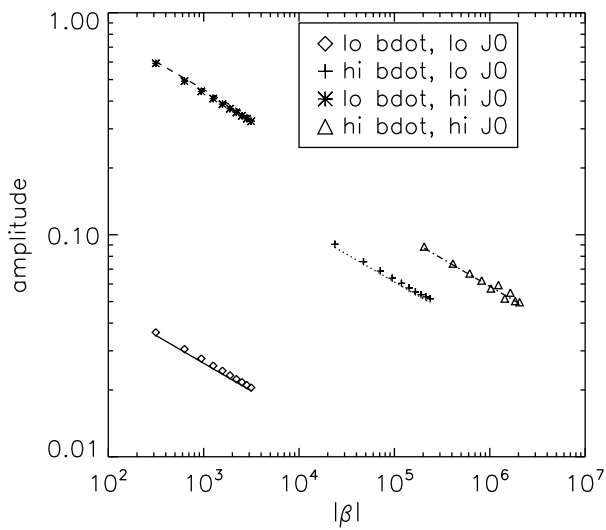}
\caption{\textbf{Top:} Evolution of libration amplitudes (in degrees) with time, for the first-order resonances. Lines show the analytical result from \S~\ref{s: evolution analytical} and points show the numerical results. 
\textbf{Bottom:} The same, for second-order resonances.}
\label{fig:width}
\end{figure}

\begin{figure}
\includegraphics[width=.5\textwidth]{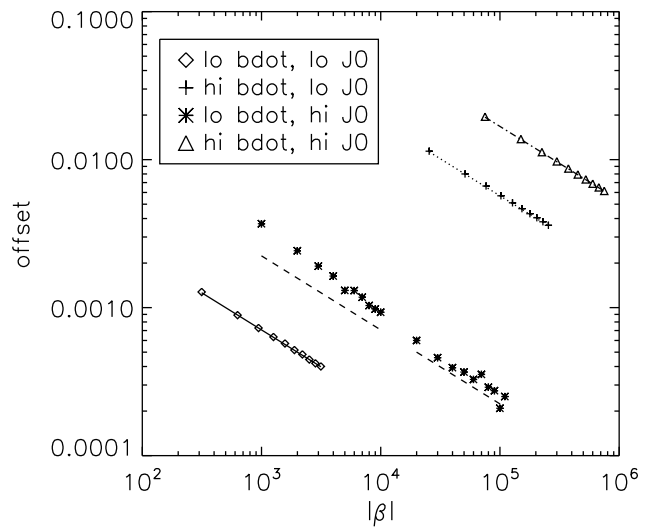}
\includegraphics[width=.5\textwidth]{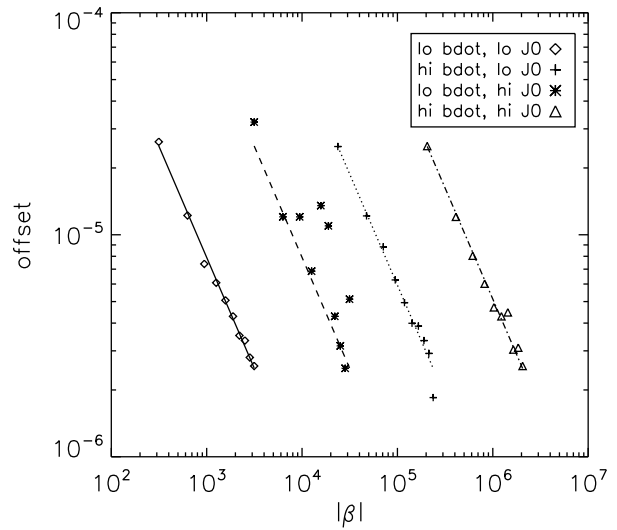}
\caption{\textbf{Top: }Evolution of the offset of libration centres (in degrees) with time, for the first-order resonances. 
\textbf{Bottom:} The same for second-order resonances}
\label{fig:centre}
\end{figure}

The analytical results in section~\ref{s: evolution analytical} show that libration amplitude and offset decrease with time, so we need to specify a time at which to measure the libration amplitude and offset. First we check the accuracy of these analytical results. For first order resonances, the top panels of Figure~\ref{fig:width} and Figure~\ref{fig:centre} show the evolution of libration amplitude and offset respectively, for one trajectory from each corner of parameter space ($J_0=0.1$, $\dot\beta=0.1$; $J_0=0.1$, $\dot\beta=2.5$; $J_0=10$, $\dot\beta=0.1$; and $J_0=10$, $\dot\beta=7.5$). The points are the numerically determined amplitudes and offsets from an extended integration of these trajectories. For the offsets, the lines are the analytical solution given by Equation~\ref{eq:centre 1st}. For the amplitudes, the lines have a slope of $-1/8$ as determined analytically, and a normalisation given by the relevant shorter integration from the grid described above. Agreement is generally good, for both fast and slow migration, even though the result was derived assuming adiabatic conservation of area. This demonstrates the accuracy of the analytical solution, and also that the output from the grid of integrations is sufficient to correctly describe the long-term behaviour.

The bottom panels of Figures~\ref{fig:width} and~\ref{fig:centre} show the analogous plots for second order resonances (trajectories taken with $J_0=0.032$,$\dot\beta=0.032$; $J_0=0.032$, $\dot\beta=2.4$; $J_0=13$, $\dot\beta=0.032$; and $J_0=13$, $\dot\beta=21$). Again we see good agreement between the analytical results and the numerical integrations.

\begin{figure}
\includegraphics[width=.5\textwidth]{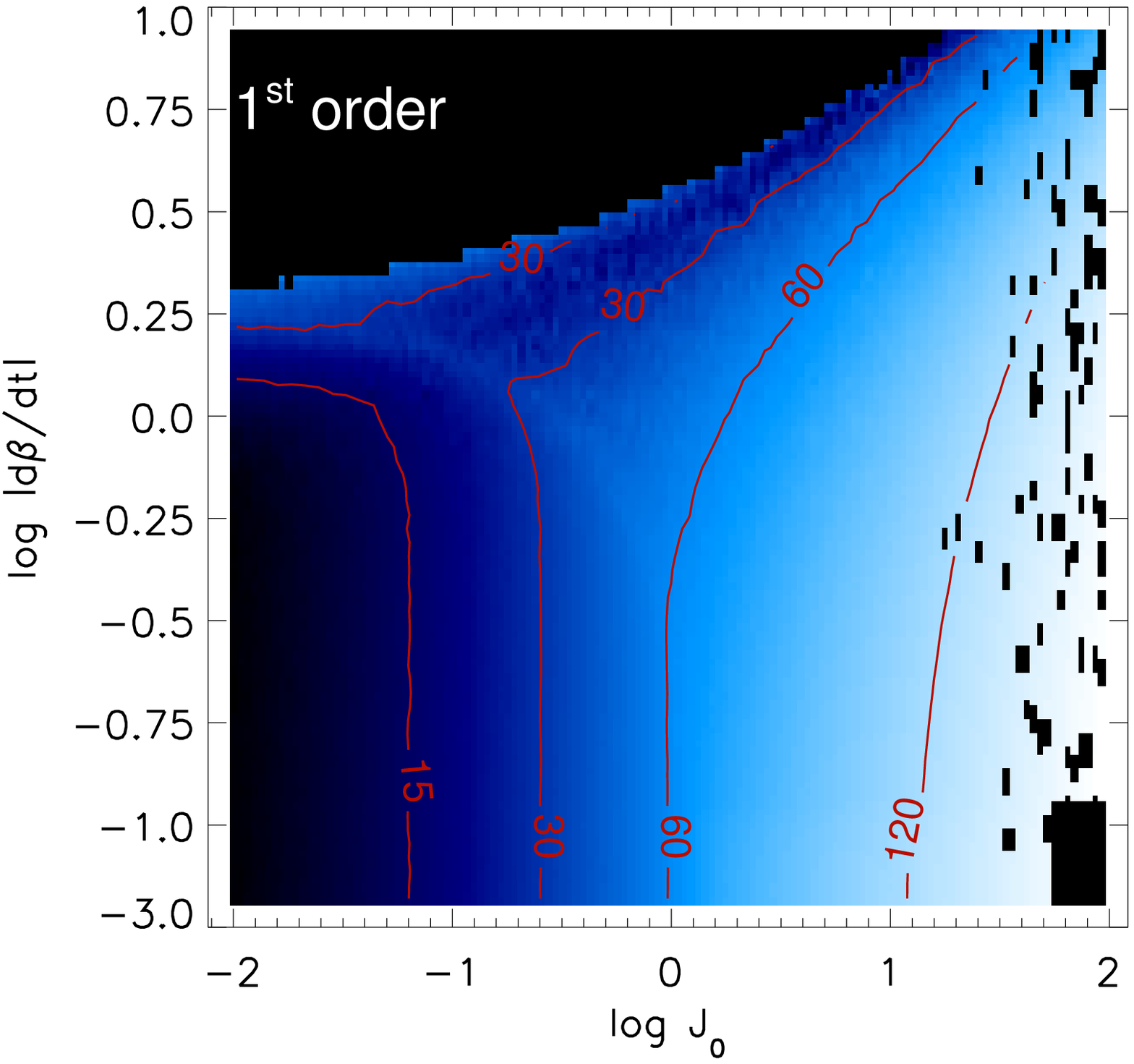}
\includegraphics[width=.5\textwidth]{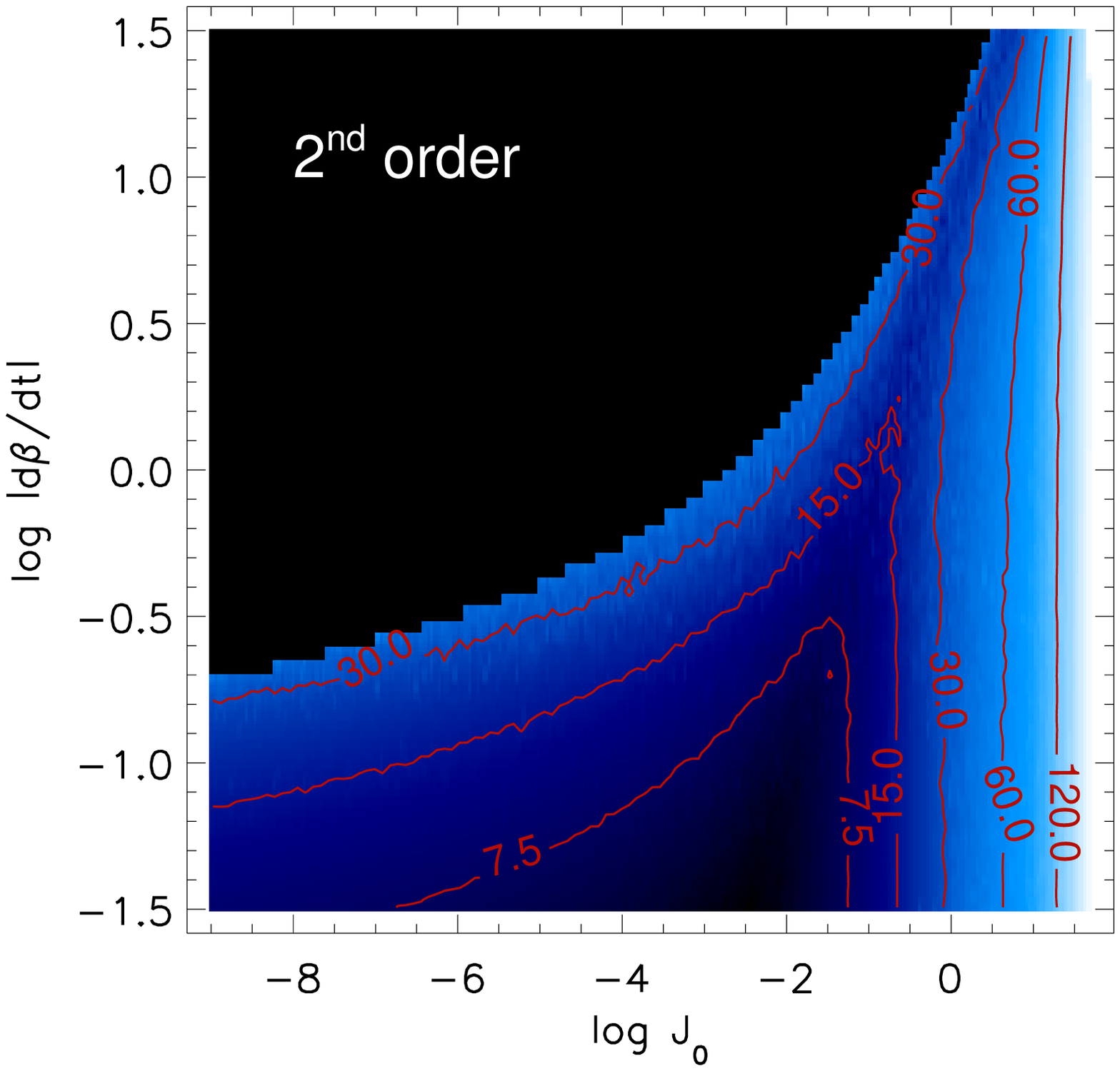}
\caption{\textbf{Top:} Mean libration amplitudes for the first-order resonances, in degrees. Amplitude is that at $\beta=-100$. Contours are at $15^\circ, 30^\circ, 60^\circ,$ and $120^\circ$. Note that the axis of ordinates is condensed at the bottom of the plot. 
\textbf{Bottom:} Mean libration amplitudes for the second-order resonances, in degrees. Amplitude is that at $\beta=-100$. Contours are at $7.5^\circ, 15^\circ,30^\circ,60^\circ,$ and $120^\circ$.}
\label{fig:amp}
\end{figure}

\begin{figure}
\includegraphics[width=.5\textwidth]{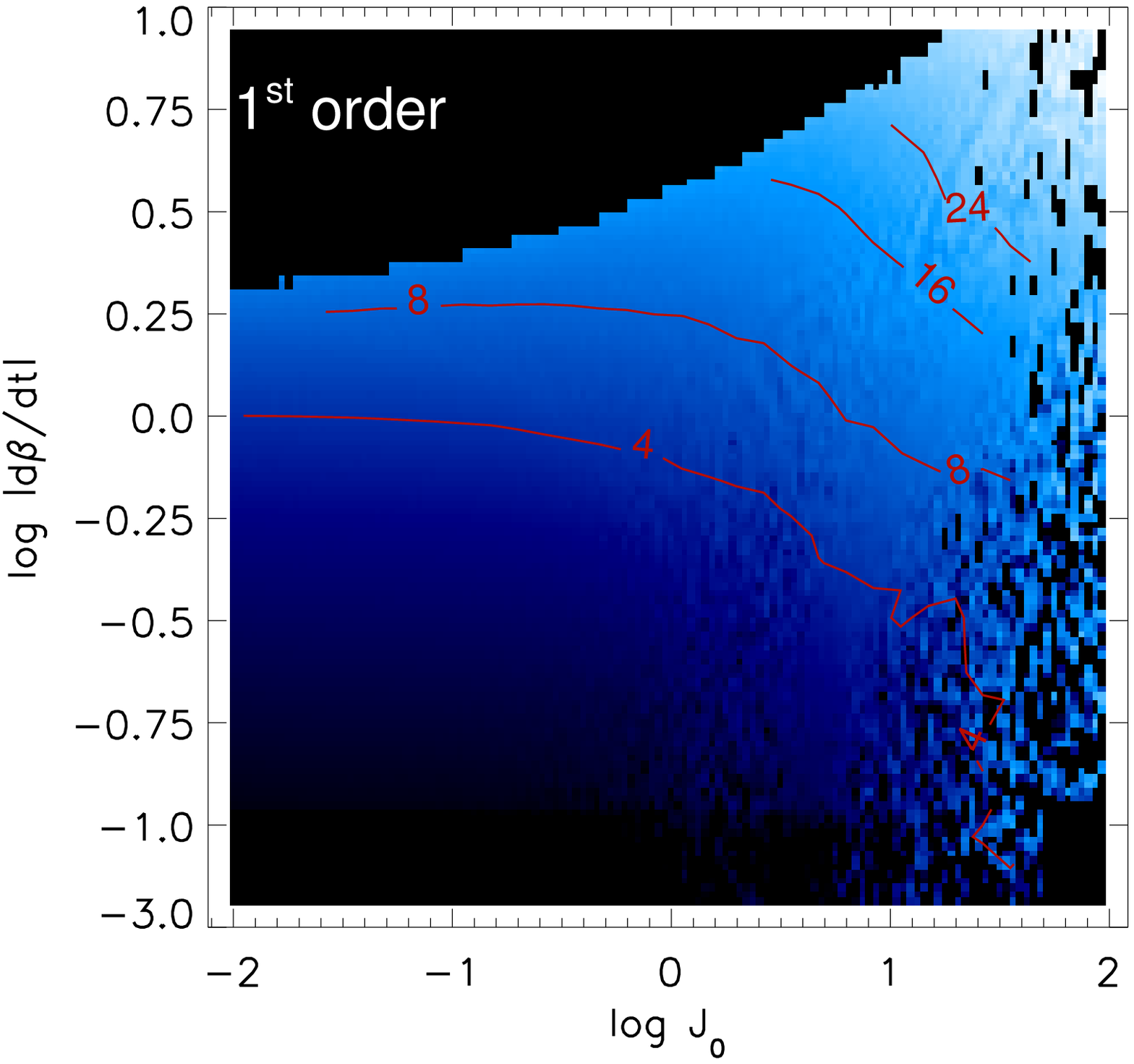}
\includegraphics[width=.5\textwidth]{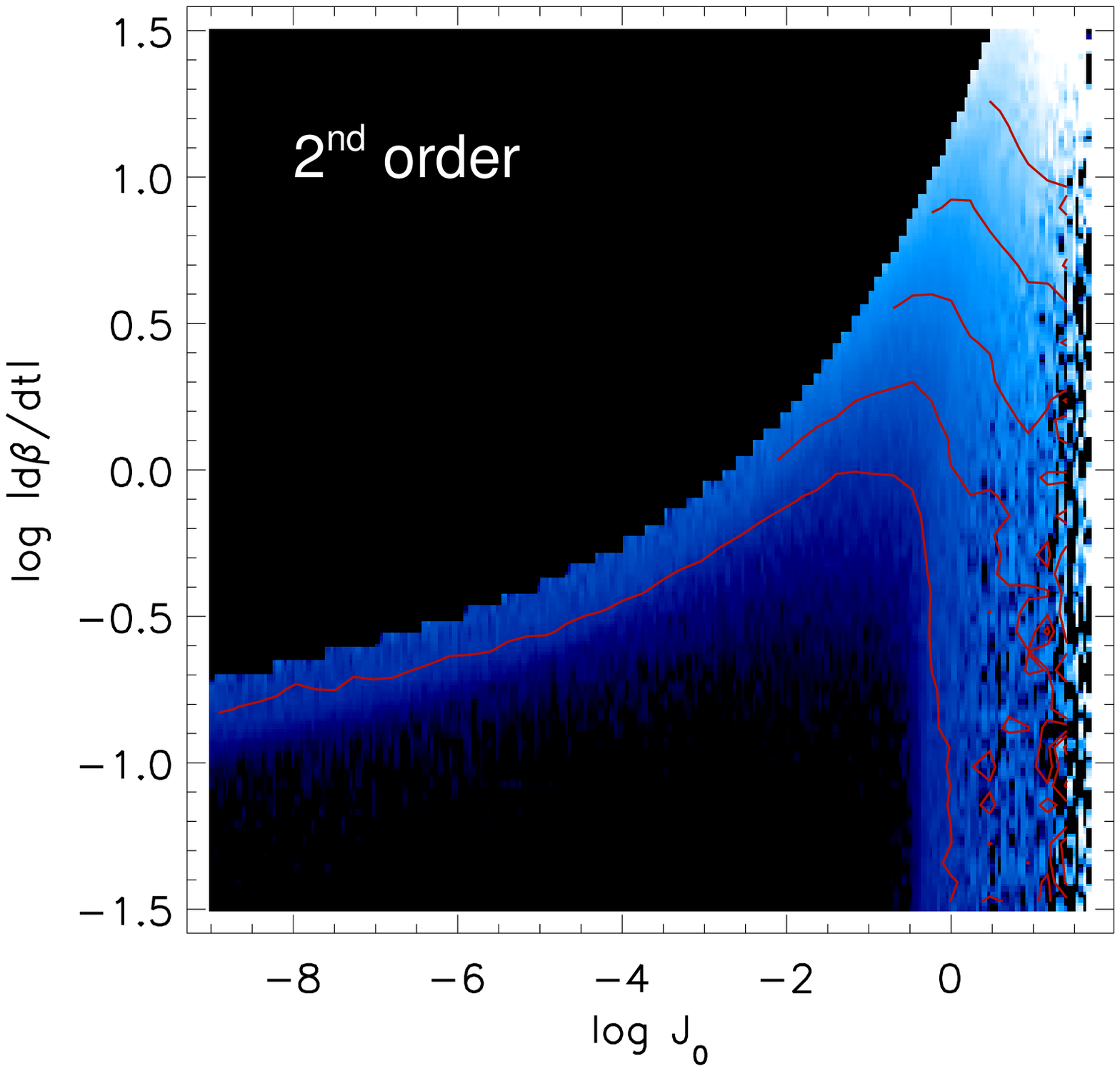}
\caption{\textbf{Top:} Mean libration offsets for the first-order resonances, in degrees. Offset is that at $\beta=-100$. Contours are at $4^\circ, 8^\circ, 16^\circ,$ and $24^\circ$. The data are noisy at high $J_0$ where there are few particles captured. Note that the axis of ordinates is condensed at the bottom of the plot. 
\textbf{Bottom:} Mean libration offsets for the second-order resonances, in degrees. Offset is that at $\beta=-100$. Contours are at $0.25^\circ$, $0.5^\circ$, $1^\circ$, $2^\circ$, and $4^\circ$.}
\label{fig:cen}
\end{figure}

The libration amplitudes and offsets for first and second order resonances for a range of $J_0$ and $\dot\beta$ are shown in figures~\ref{fig:amp} and~\ref{fig:cen}. Note that all amplitudes and offsets are scaled to what they would be at $\beta=-100$ using the time dependence given in equations \ref{eq:centre 1st}, \ref{eq:libwidth 1st}, \ref{eq:centre 2nd}, and \ref{eq:amp 2nd}. Large amplitude librations can occur as a result of either large initial momentum or, to a lesser extent, faster migration. For first order resonances, at low momentum the libration offset is not dependent on $J_0$ and increases with migration rate, as predicted by analytic theory \citep{1994Natur.369..719D}. At higher momenta the offset decreases with increasing $J_0$, although here the data are noisy due to low numbers of particles being captured. For second order resonances the offset increases with increasing $\dot\beta$, and as $J_0$ increases the offset decreases, passing through a minimum before increasing again at high $J_0$.

Note that the offset is only present during migration: if the migration force is removed, such as if a perturbing gas disc has dissipated, then libration will be about $\pi$ exactly with no offset.

\subsection{Failure to capture}

\begin{figure}
\includegraphics[width=.5\textwidth]{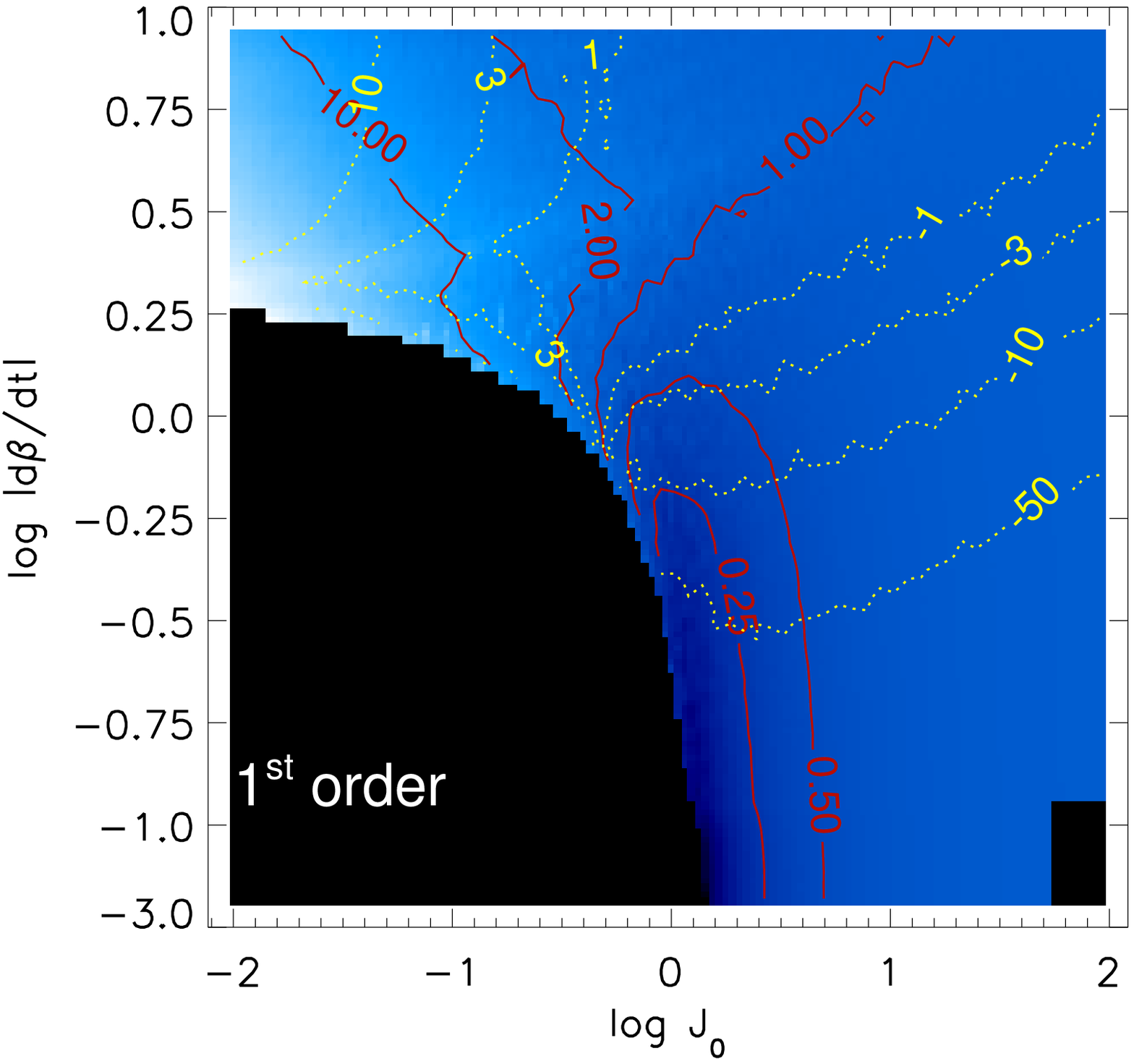}
\includegraphics[width=.5\textwidth]{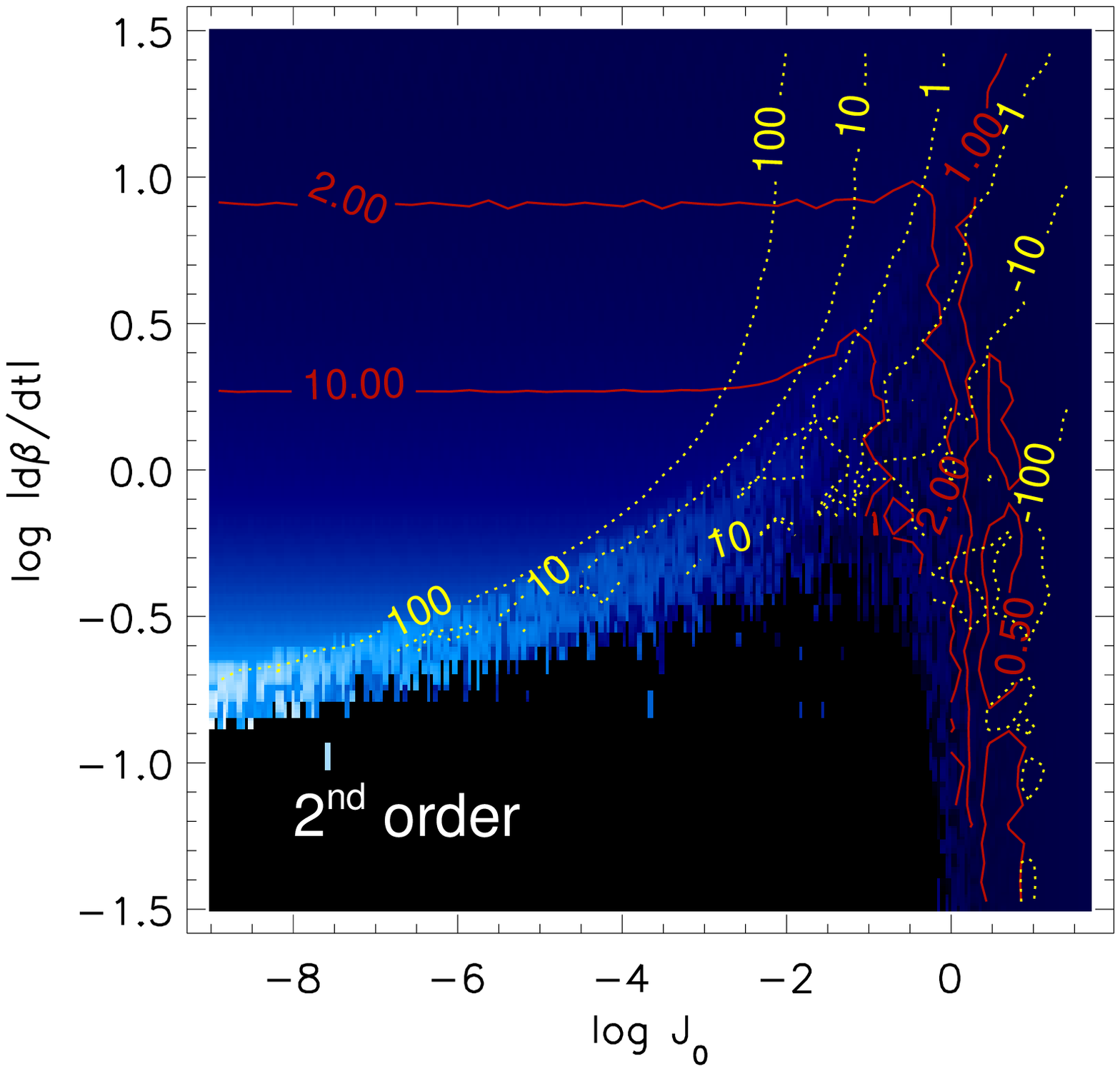}
\caption{\textbf{Top:} Change in eccentricity when passing through a first-order resonance, in the event of a non-capture. Colour gradient and associated solid contours show the mean final momentum $J$ as a fraction of the initial $J_0$, so that particles on the 1.00 contour experience no mean change, particles with ratio greater than 1.00 are pumped up on average, and particles with ratio less than 1.00 are cooled on average. The dotted contours show the mean change divided by the standard deviation. At low migration rates the distribution is narrow and all particles behave in essentially the same way---they are cooled---but at high migration rates there is a large spread in eccentricity jumps. 
\textbf{Bottom:} The same, for second order resonances.}
\label{fig:jump}
\end{figure}

In the event that the particle not be trapped in resonance, its eccentricity changes as it passes through the resonance. In the case of adiabatic migration, the eccentricity is reduced \citep{1999ssd..book.....M}. On the other hand, we find that when the particle fails to capture due to fast migration rate, the eccentricity can change substantially in either direction. The top panel of Figure~\ref{fig:jump} illustrates how the eccentricities are changed when passing through a first-order resonance. The colour scale and solid contours show the final eccentricity as a fraction of the initial eccentricity. We see agreement with adiabatic theory in the case of slow migration: all particles' eccentricities are driven down. In Figure~\ref{fig:jump} we also show contours of (mean change in momentum)/(standard deviation of change in momentum) to illustrate how broad the distribution of eccentricity changes is relative to the mean. For slow migration rates the distribution is very narrow, and all particles behave in the same manner. In contrast, for faster migrations and high momentum we see that not only can the average change in eccentricity be positive, but that the distribution of changes is relatively broad, so that many particles lose eccentricity despite the mean being an increase. In this case, the behaviour is highly stochastic. For fast migration and low momentum, the particles' eccentricities are all driven up. We note furthermore that the actual distribution of eccentricity jumps is highly non-Gaussian, being either unimodal and strongly skewed or even bimodal. Note that for low initial eccentricities the mean eccentricity jump decreases as the migration rate is increased, presumably because the resonance has less time to affect the particle's orbit before being crossed. 

The bottom panel of Figure~\ref{fig:jump} shows an analogous plot for the second-order resonances. The behaviour is qualitatively similar: with slow migration the eccentricity of all particles is decreased on passing through the resonance, but for fast migration and low eccentricity the eccentricity is pumped up, while at higher eccentricity the eccentricity may jump either up or down.

\section{Model validation}

\begin{figure}
\includegraphics[width=.5\textwidth]{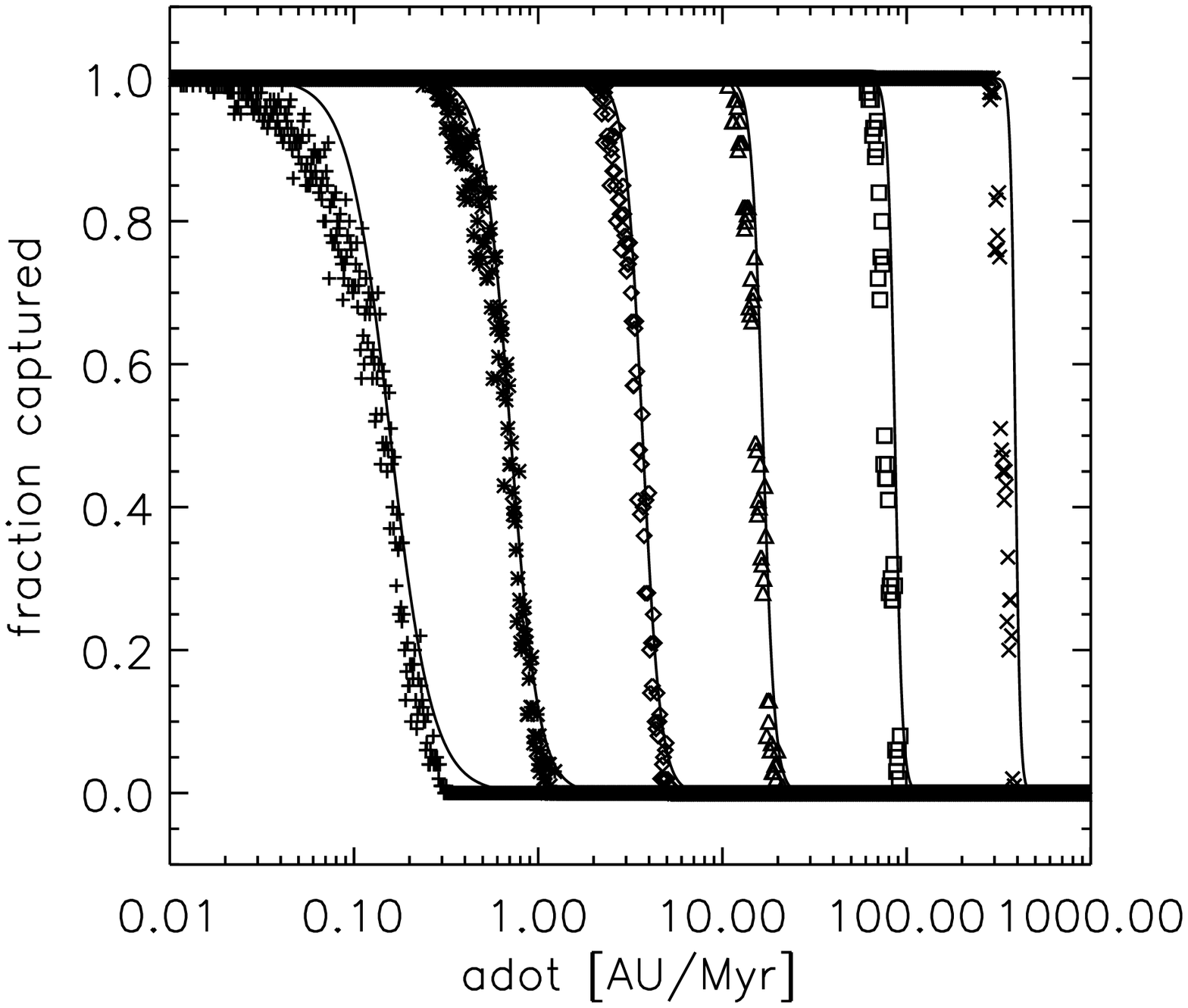}
\includegraphics[width=.5\textwidth]{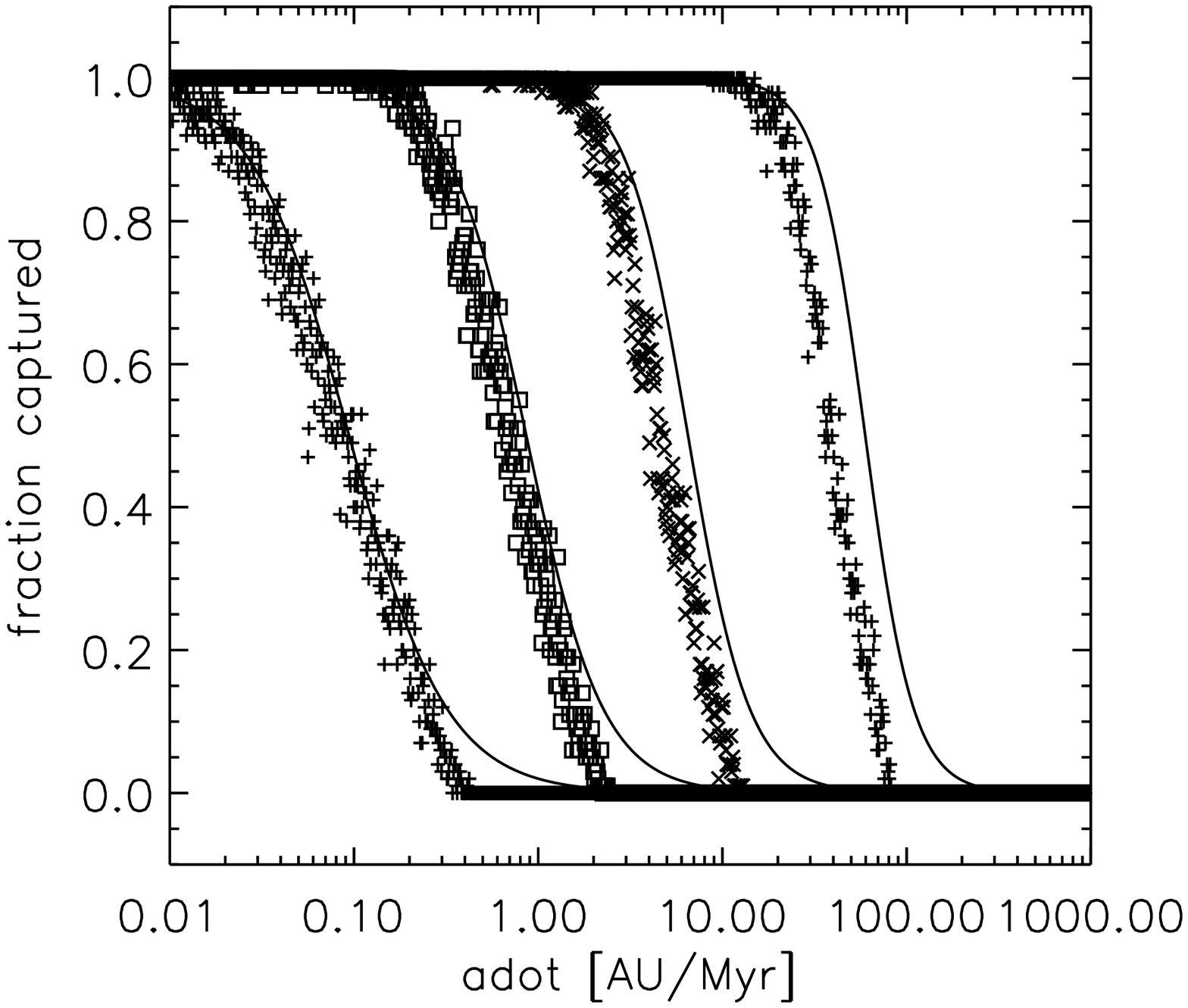}
\caption{\textbf{Top:} Capture probabilities as a function of planetary migration rate for the 3:2 external resonance. Solid lines are the fitting formulae from the N-body simulations of W03, and points are from the Hamiltonian model. From left to right, planet masses are 1, 3, 10, 30, 100 and 300 times Earth's mass. 
\textbf{Bottom:} Capture probabilities as a function of planetary migration rate for the 5:3 external resonance. Solid lines are the fitting formulae from W03, and points are from the Hamiltonian model. From left to right, planet masses are 30, 100, 300 and 1000 Earth's mass.}
\label{fig:wyatt03 prob}
\end{figure}

We compared line cuts through our figures of capture probability to figures~2 and 3 of Q06, demonstrating that our integrations of the Hamiltonian are accurate. To demonstrate that the Hamiltonian model itself correctly describes the behaviour of the three body problem, we now compare our results with the N-body simulations of W03, who considered the case where a massive planet was migrating outward into a disc of test particles. We conduct a Monte-Carlo simulation, creating a population of particles with the same distribution as in the N-body integration and converting their parameters into the dimensionless ones to determine the outcome of the resonance passage process. This process is very quick: each point in Figure~\ref{fig:wyatt03 prob} represents 100 particles, so there are around $10^6$ particles in total, and the process took only a few seconds. Having generated our samples we plot their capture probabilities and amplitudes and offsets against migration rate, and also show the empirical fitting formulae from Wyatt's N-body simulations.

The top panel of Figure~\ref{fig:wyatt03 prob} shows the capture probability as a function of planetary migration rate for the 3:2 external resonance. Test particles were located at 60\,AU, with eccentricities in the range [0,0.01]. Hence the population for each point in $(\dot{a}, m_\mathrm{pl})$ space is taken form a horizontal cut through the dimensionless parameter space, up to a certain maximum value of $J$. The stellar mass is 2.5 Solar masses and the planet mass ranges from 1 to 300 Earth masses. The probabilities obtained from the Hamiltonian model show excellent agreement with those obtained from N-body integrations (see figure 1 of W03); in particular we reproduce the increasing sharpness of the transition from certain capture to certain failure to capture as the planet mass is increased. This is because for low mass planets, an eccentricity of only 0.01 is sufficiently non-zero to affect the trapping probability, putting the particles in the region of parameter space where the transition from certain capture to impossible capture as migration rate increases begins to broaden. Models involving capture into resonance with low mass planets will therefore need to pay careful attention to the initial conditions of the planetesimal disc, since the distribution of eccentricities, even in a dynamically cold disc, can affect resonant trapping behaviour. In contrast, capture into resonance with high mass planets will follow the same behaviour for all small eccentricities.

The capture probabilities for second-order resonances, too, show excellent agreement with N-body simulations. Compare the bottom panel of Figure~\ref{fig:wyatt03 prob} with figure~4c of W03, for the case of the 5:3 resonance. The second order resonances are much weaker than the first order ones, and we cover the same range of migration rates with more massive planets: from 30 to 1000 earth masses.

\begin{figure}
\includegraphics[width=.5\textwidth]{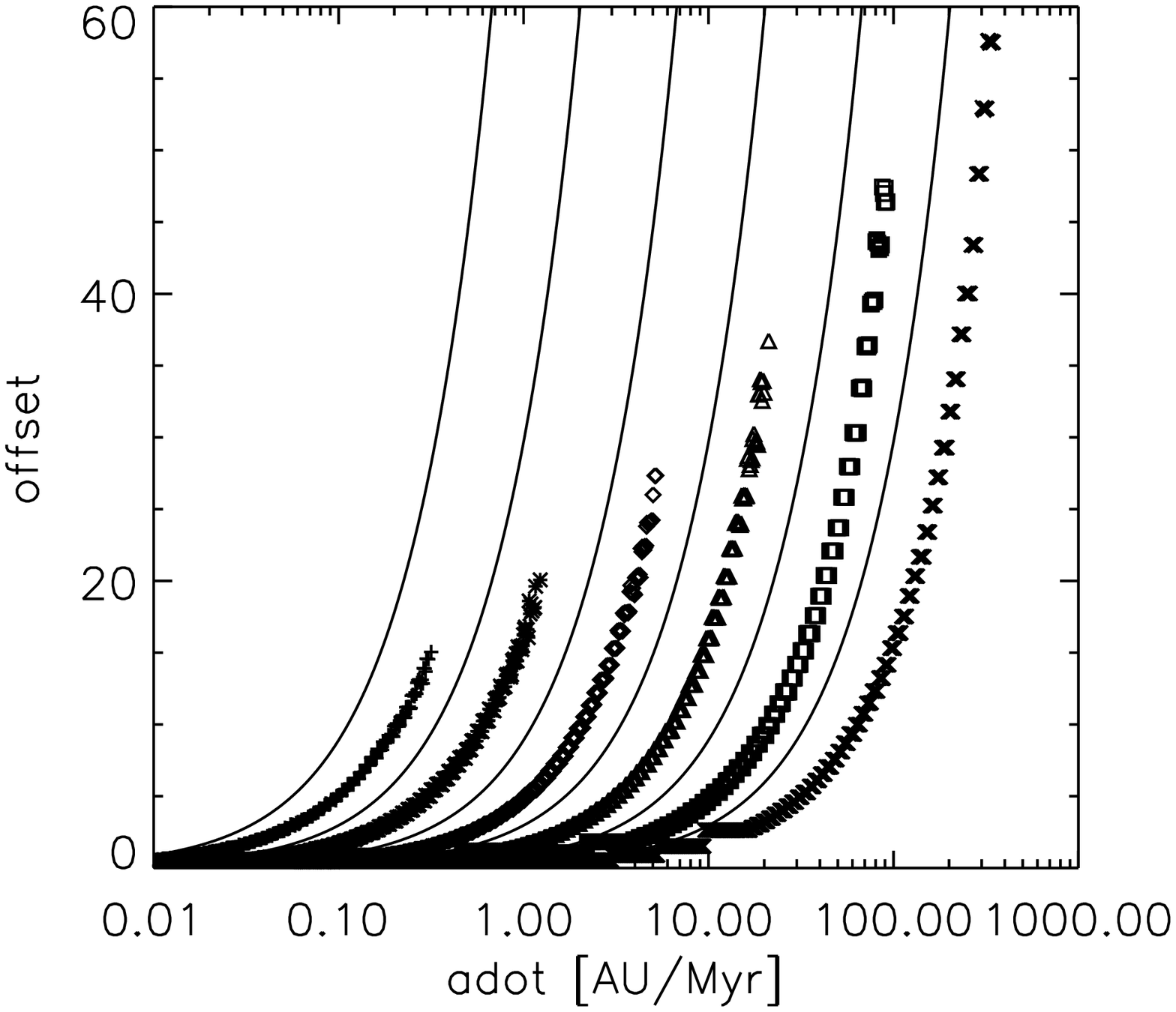}
\includegraphics[width=.5\textwidth]{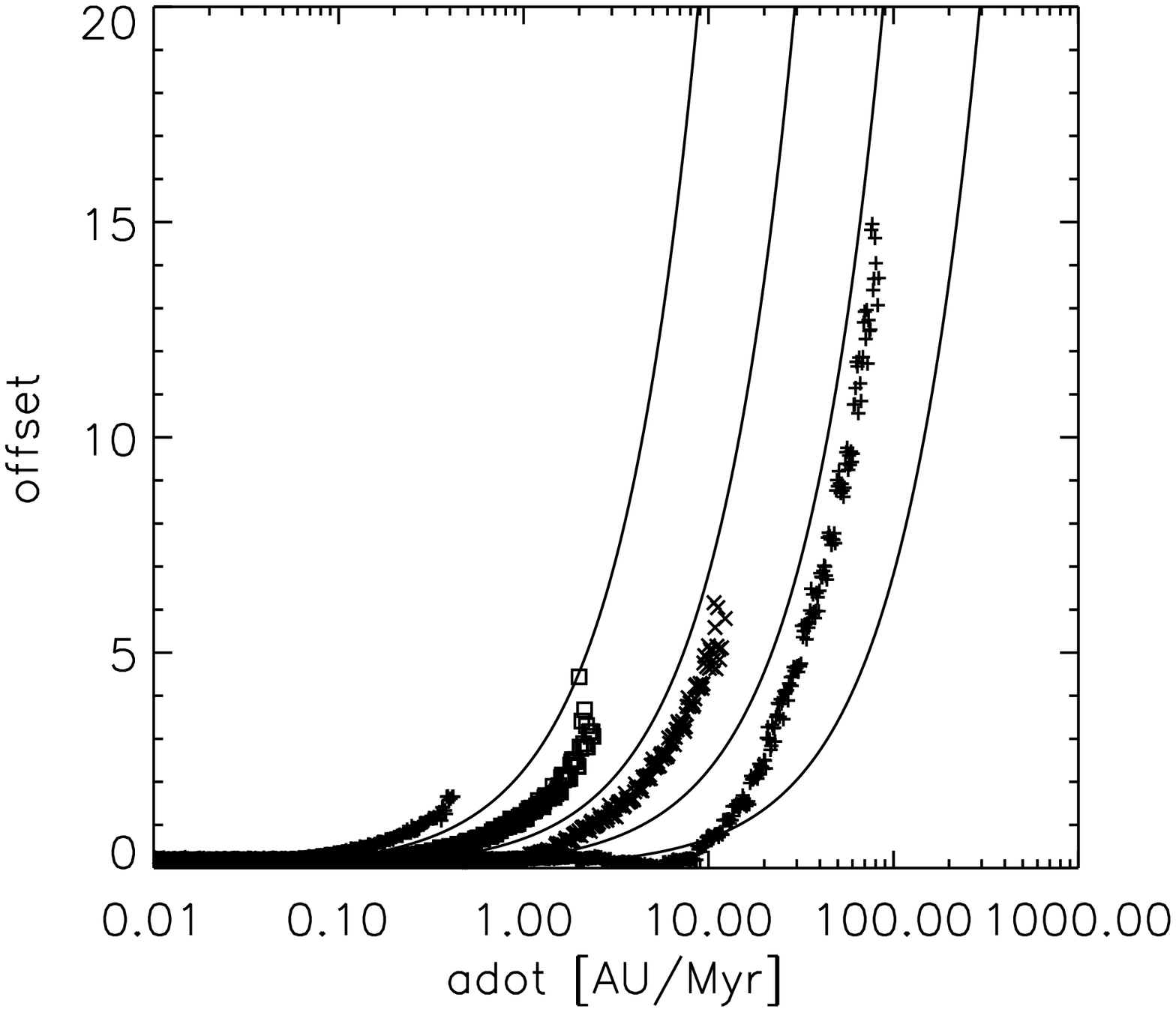}
\caption{\textbf{Top:} Libration offsets for the 3:2 external resonance. Solid lines are the fitting formulae from W03, and points are from the Hamiltonian model. From left to right, planet masses are 1, 3, 10, 30, 100 and 300 times Earth's mass. 
\textbf{Bottom:} Libration offsets for the 5:3 external resonance. Solid lines are the fitting formulae from W03, and points are from the Hamiltonian model. From left to right, planet masses are 30, 100, 300 and 1000 times Earth's mass.}
\label{fig:wyatt03 offset}
\end{figure}

\begin{figure}
\includegraphics[width=.5\textwidth]{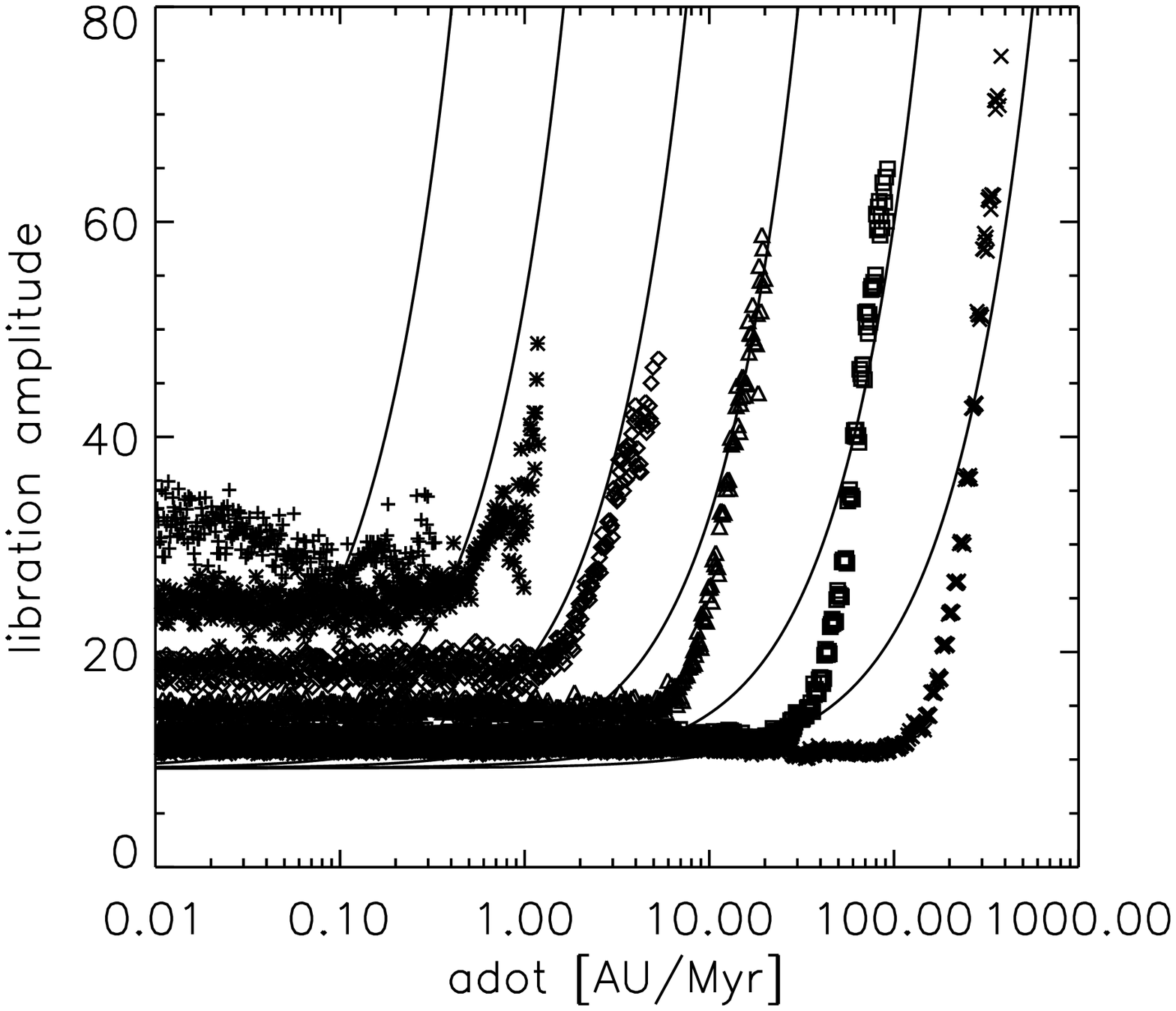}
\includegraphics[width=.5\textwidth]{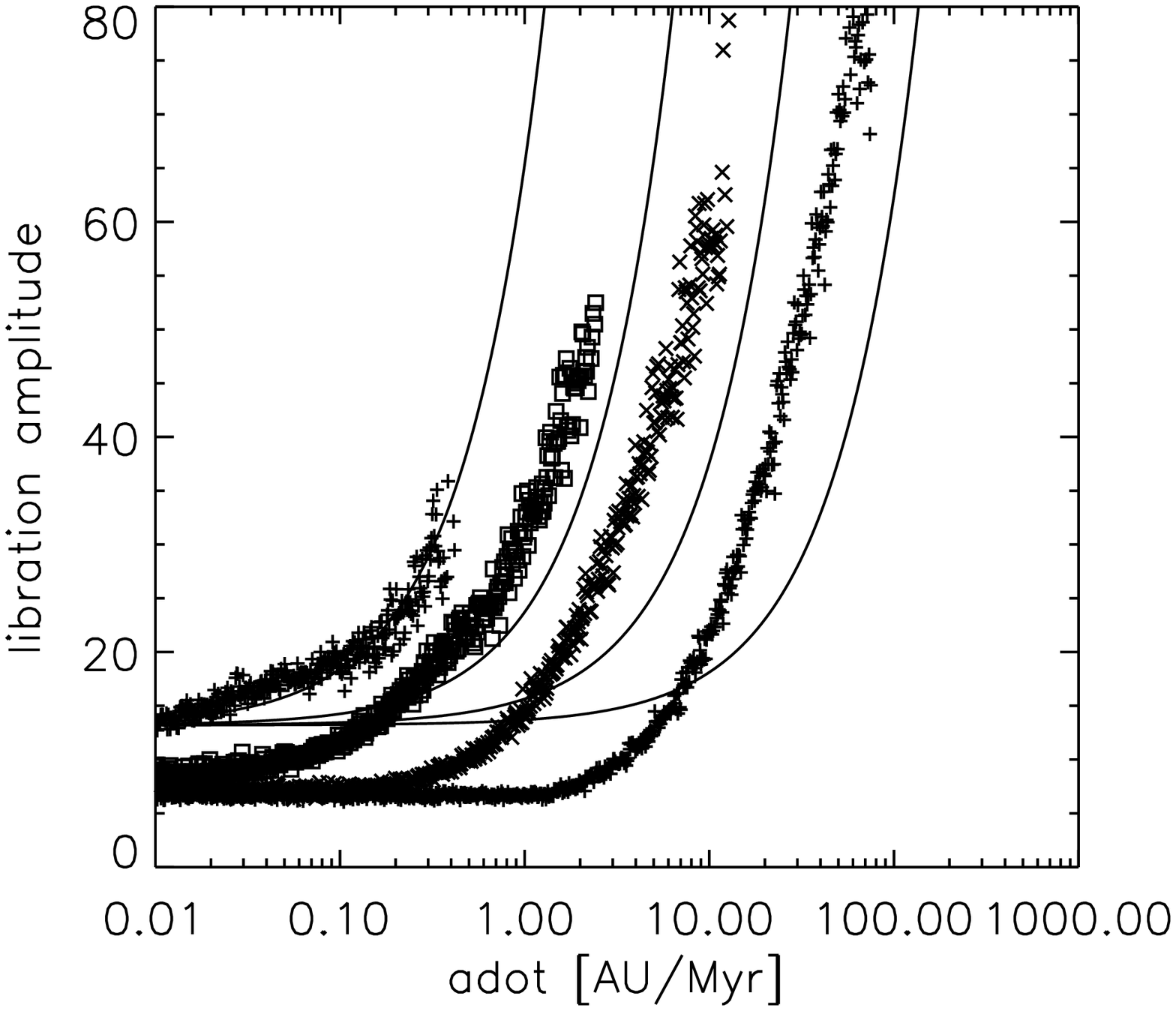}
\caption{\textbf{Top:} Libration amplitudes for the 3:2 external resonance. Solid lines are the fitting formulae from W03, and points are from the Hamiltonian model. From left to right, planet masses are 1, 3, 10, 30, 100 and 300 times Earth's mass. 
\textbf{Bottom:} Libration amplitudes for the 5:3 external resonance. Solid lines are the fitting formulae from W03, and points are from the Hamiltonian model. From left to right, planet masses are 30, 100, 300 and 1000 times Earth's mass.}
\label{fig:wyatt03 width}
\end{figure}

In the top panels of figures~\ref{fig:wyatt03 offset} and \ref{fig:wyatt03 width} we compare the offsets and amplitudes from the Hamiltonian model with the results from \cite{2003ApJ...598.1321W}, for the 3:2 external resonance. We see qualitative agreement between the Hamiltonian model and the N-body fits, although quantitative agreement is not exact. We see the libration offset steadily increasing with migration rate, although that from the model is somewhat lower. The libration amplitudes are, for planets more massive than $1\mathrm{\,M}_\oplus$, constant for small migration rates before increasing at higher migration rates, although for Earth mass planets the amplitude is independent of migration rate. At a given migration rate, capture by a less massive planet will result in a larger libration amplitude. The fitting formulae from W03 shown on this plot were found by aggregating over all planet masses, so the Hamiltonian model provides a more accurate prediction for a given mass. The difference between the Hamiltonian and N-body results may be due to W03 not using a consistent duration of migration post capture; we have taken the migration to continue for 0.5\,AU in all cases. We also have a semi-analytical motivation for Wyatt's finding that the libration amplitude depends on $\dot{a}m_\mathrm{pl}^{-1.3}$: the libration amplitude depends on the dimensionless migration rate $\dot\beta$, which is in turn proportional to $\dot{a}m_\mathrm{pl}^{-4/3}$ (equation~\ref{eq:scalings first}).

The bottom panels of Figures~\ref{fig:wyatt03 offset} and \ref{fig:wyatt03 width} show the analogous plots for the 5:3 resonance. Again the model reproduces the qualitative behaviour of the N-body results, with the model this time yielding slightly higher offsets, and similar differences between the model and N-body as in the case of first order resonances. Again, higher libration amplitudes result from lower planet masses at a given migration rate.

\section{Discussion}

\subsection{General comments}

\label{S:general}

To summarize the previous sections, we have shown that capture into a resonance is certain at low eccentricities and migration rates, impossible at fast migration rates and low eccentricities, and possible but not certain for high eccentricities. The critical eccentricity at which capture becomes probabilistic is lower for lower planet masses: around 0.1 for a Jupiter mass planet and around 0.01 for an earth mass planet, for first order resonances. Furthermore, at lower eccentricity, the width of the transition from capture to no capture with increasing $\dot{a}$ is also affected. The effects of this can be seen in Figure~\ref{fig:wyatt03 prob}. Eccentricities of order 0.1 may not be uncommon in planetesimal belts: bodies in the classical Kuiper Belt currently have a mean eccentricity of about 0.1 \citep[][]{2002ARA&A..40...63L}, and eccentricities in planetesimal belts can be pumped up above 0.1 when the largest bodies grow significantly and gravitationally perturb their neighbours \citep{2008ApJS..179..451K}. Thus, the role of eccentricity cannot be neglected when investigating resonant encounters. If capture into a resonance occurs, the resulting libration amplitude is larger if initial eccentricity was larger, and, for small initial eccentricities, also increases with increasing migration rate. Libration centre is offset by an amount dependent only on migration rate, in agreement with analytic theory, unless the dimensionless momentum is very high. If a particle fails to be captured, the eccentricity is invariably driven down for slow migration rates, but can jump up or down for high migration rates.

Before discussing specific scenarios, we make some general comments.

\begin{figure}
\includegraphics[width=.5\textwidth]{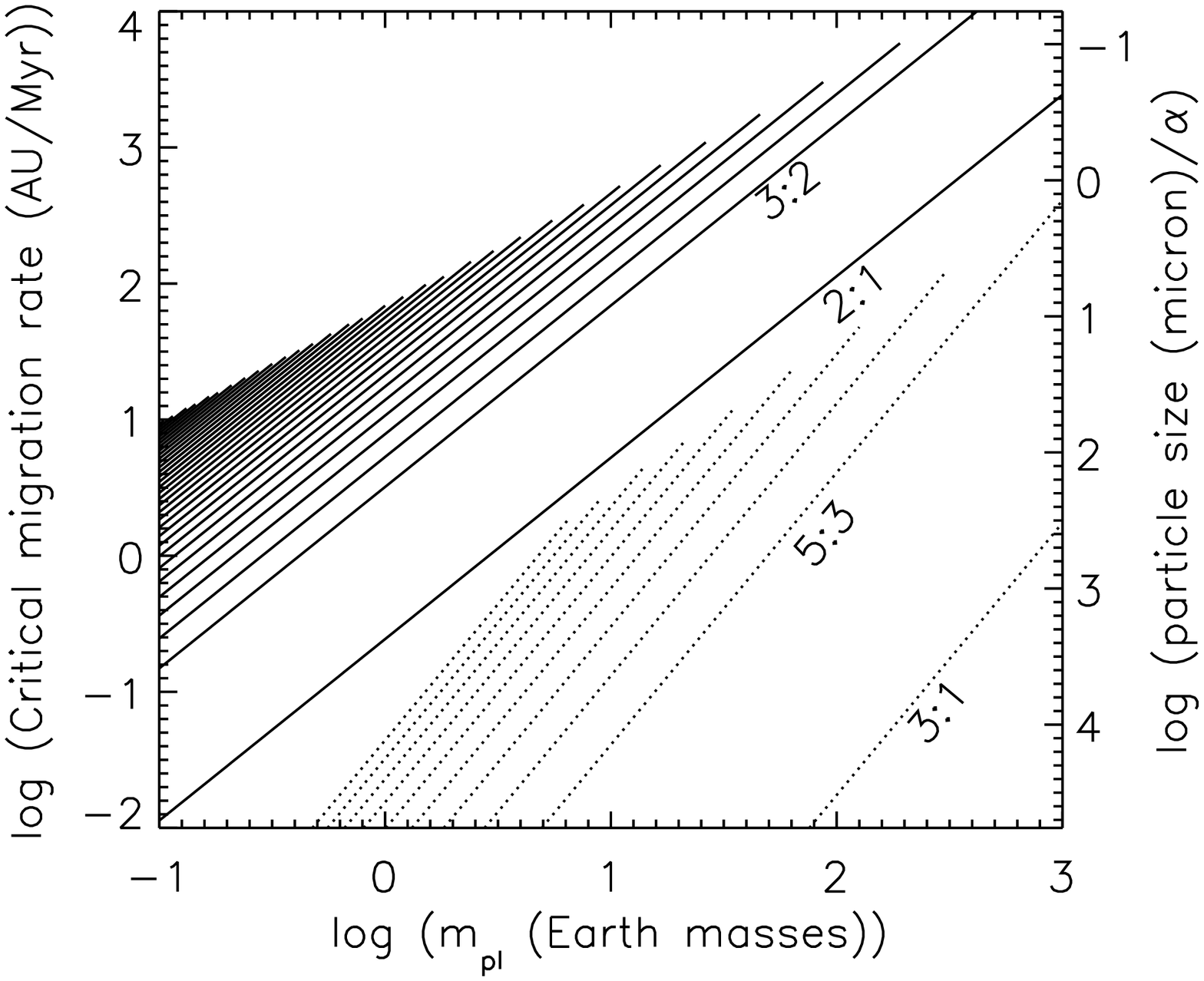}
\includegraphics[width=.5\textwidth]{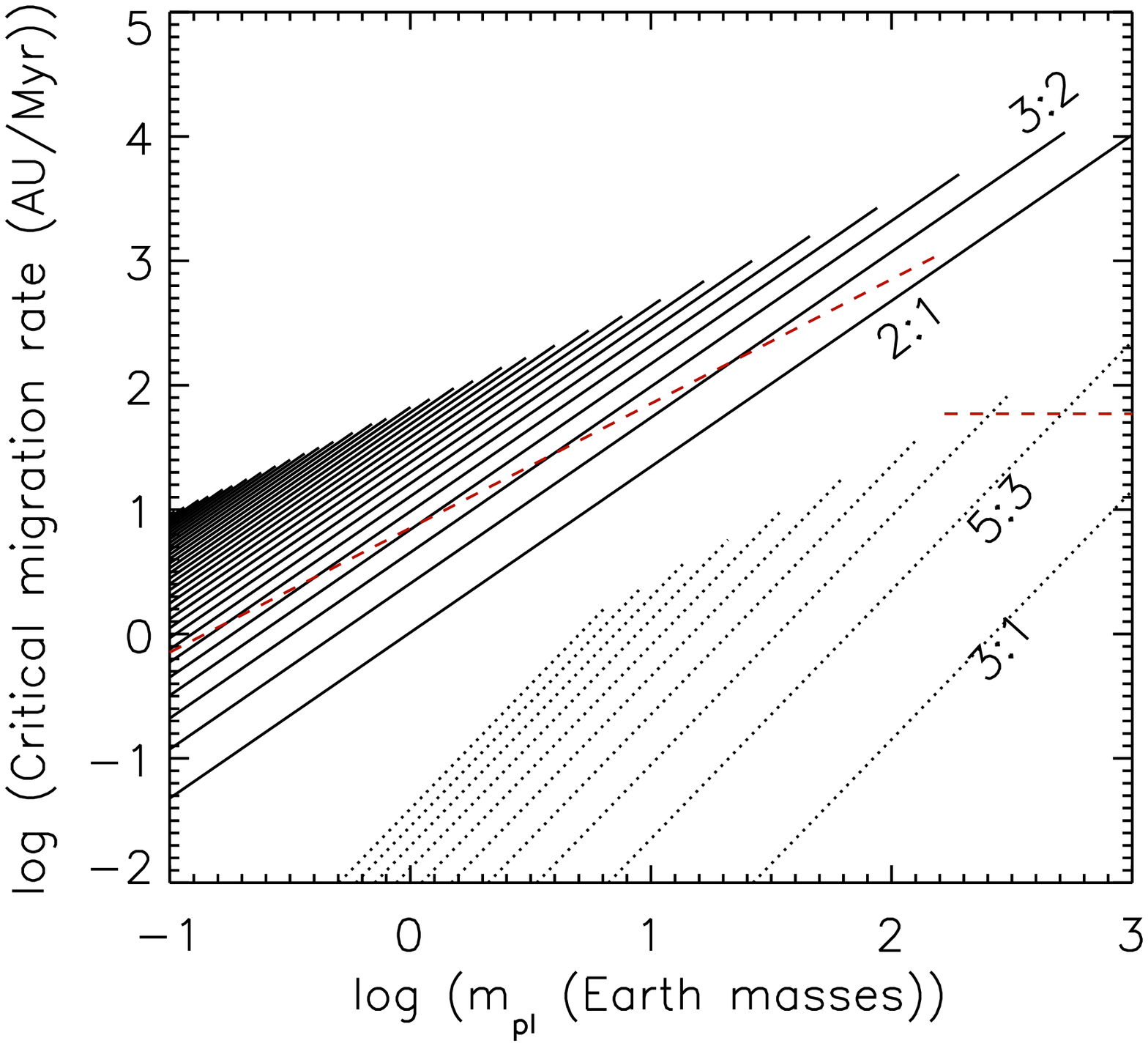}
\caption{\textbf{Top:} Critical migration rate at 1\,AU for capture of low-eccentricity particles drifting into resonances exterior to a planet as a function of planet mass. Stellar mass is Solar. At migration rates faster than this, capture is impossible at low eccentricity. The critical migration rates for first order resonances are shown as solid lines. The 2:1 and 3:2 resonances are labelled; the higher $j$ resonances move away from these monotonically. The line for a given resonance is terminated when that resonance becomes unstable according to the resonance overlap criterion. We also show second-order resonances as dotted lines; only up to $j=21$ are shown for clarity ($\dot\beta=0.3$ was taken as the critical migration rate, although this depends on eccentricity; see Figure~\ref{fig:prob}. $\dot\beta=0.3$ is the maximum of the envelope of the 100\% capture region.). The right-hand axis of ordinates shows the equivalent dust grain size for migration under PR drag (see \S\,\ref{S:zodi}).  
\textbf{Bottom:} The same, but for trapping of particles interior to a migrating planet. We show typical migration rates for planets embedded in gas discs at 1\,AU (dashed lines). Mass-dependent Type I migration occurs for low mass planets and mass-independent Type II migration for high mass planets; see \S\,\ref{S:planets}}
\label{fig:adotcrit}
\end{figure}

For any migration scenario, if migration lasts sufficiently long, a succession of resonances will be encountered, if capture fails to occur at the first resonance. For low eccentricities, the critical migration rate for capture increases with $j$, so a particle failing to capture into the 2:1 resonance may be captured into the 3:2, 4:3, etc. Indeed, the critical migration rate increases without bound as $j\to\infty$, so it would appear that ultimately a particle will be captured into a resonance very close to the planet. However, such resonances are not stable. Close to the planet the resonances overlap, forming an unstable region of chaotic behaviour of width $\approx 1.4a_\mathrm{pl}(m_\mathrm{pl}/m_\star)^{2/7}$ \citep{1980AJ.....85.1122W}. Thus the closest first-order resonance which is stable is given by $j_{\max}=18(m_\mathrm{pl}/m_\oplus)^{-2/7}(m_\star/m_\odot)^{2/7}$. Even if particles were captured into these resonances, they would be quickly removed. Hence for a given planet mass there is a maximum $j$ which will trap particles, and conversely, for a given resonance there is a maximum planet mass, which is determined by the width of the chaotic zone. This is illustrated in Figure~\ref{fig:adotcrit}. The top and bottom panels are for capture of particles exterior to and interior to the planet, respectively. The diagonal lines show the critical migration rate for capture into resonances of a given $j$ as a function of planet mass. This figure generalises figure~5 of W03 to resonances of higher $j$, illustrating the power of the Hamiltonian model. 

The lines in Figure~\ref{fig:adotcrit} are terminated by the chaotic zone prescription described above. For example, a $1\mathrm{\,M}_\oplus$ planet at 1\,AU will be able to capture exterior low eccentricity particles in resonances up to the 18:17. To calculate the upper envelope, we know that the critical dimensionless migration rate is $\dot\beta\approx 2.1$ (\S\,\ref{capture first order}). From table~\ref{tab:l_j}, we find a fit $l_j\approx 5.4j^{-1.8}$, so the relevant coefficient for the innermost stable resonance is $l_{j,\max}\approx 0.0297(m_\mathrm{pl}/m_\oplus)^{0.514}(m_\star/m_\odot)^{-0.514}$. Using Equation \ref{eq:scalings first}, we therefore have
\begin{eqnarray}\label{eq:acrit any ext}
\left[\alpha\frac{\dot{a}_2}{\mathrm{AU\,Myr}^{-1}}-\frac{\dot{a}_1}{\mathrm{AU\,Myr}^{-1}}\right]_\mathrm{crit}
&\approx&70\left(\frac{m_\mathrm{pl}}{m_\oplus}\right)^{0.8}
\left(\frac{m_\star}{m_\odot}\right)^{-0.3}\nonumber\\
&&\left(\frac{a_\mathrm{pl}}{\mathrm{AU}}\right)^{-1/2},
\end{eqnarray}
above which capture is impossible into any resonance, unless the eccentricity is sufficiently high. This equation holds for both external and internal particles.

While Figure~\ref{fig:adotcrit} is plotted for a planet at 1\,AU, similar results can be obtained for other distances: the critical migration rate for a given resonance decreases ($\propto 1/\sqrt{a_\mathrm{pl}}$, from equation~\ref{eq:scalings first}) and the critical migration rate for capture into any stable resonance has the same dependence. Thus, slightly lower migration rates are required for planets further from the star to capture particles than planets of the same mass in smaller orbits. The effect of this will depend on the migration mechanism: for Poynting--Robertson drag, the migration rate is inversely proportional to $a$, so lower $j$ resonances will capture particles when the planet is further from the star. On the other hand, for Type I migration in a disc with a MMSN power law index, the migration rate is independent of $a$, so capture will be into higher $j$ resonances if the planet is further from the star.

\begin{figure}
\includegraphics[width=.5\textwidth]{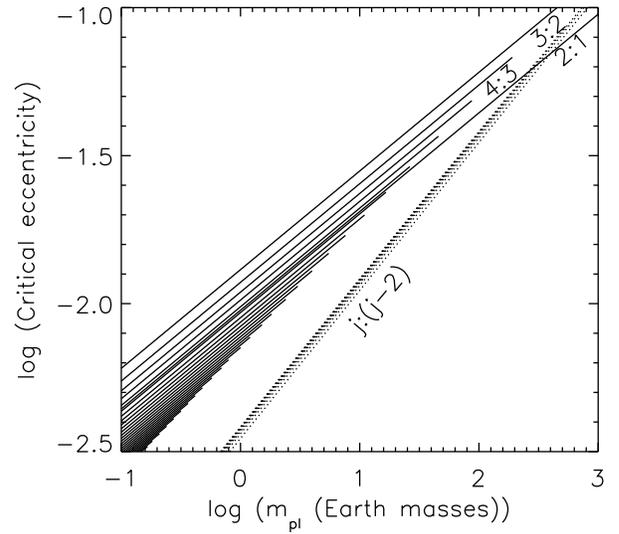}
\caption{Critical eccentricity for certain capture of slowly migrating particles into external resonances as a function of planet mass. The star is of Solar mass and the planet is located at 1\,AU. At eccentricities greater than this capture is possible but becomes less probable with increasing $e$. Critical eccentricities for first order resonances are shown with solid lines. The 2:1, 3:2, and 4:3 resonances are labelled; the higher $j$ resonances move away from the 3:2 and 4:3 monotonically. The line for a given resonance is terminated when that resonance becomes unstable according to the resonance overlap criterion. Critical eccentricities for second order resonances are shown as dotted lines.}
\label{fig:ecrit}
\end{figure}

We have ignored the effects of eccentricity in the previous paragraph. This plays an ambiguous role: at low migration rates eccentricity reduces the probability of capture. For first-order resonances, the critical eccentricity for the transition from certain to probabilistic capture decreases with increasing $j$ (Figure~\ref{fig:ecrit}). For second-order resonances the critical eccentricity does not depend strongly on $j$.  However, at high migration rates, higher eccentricity can make capture possible where capture was impossible at low migration rates; this effect is particularly noticeable for second-order resonances. Furthermore, resonances can cause particles' eccentricities to change as the particles pass through resonance and fail to be captured, and the cause of migration itself often damps eccentricities too \citep[][although disc--planet torques may also pump eccentricity, e.g., \citealt{2008Icar..193..475M}]{1979Icar...40....1B,2004ApJ...602..388T}.

If a particle encounters a succession of resonances, then, ignoring changes in eccentricity due to resonant encounters or other effects, the dimensionless migration rate will decrease and the dimensionless momentum increase at each subsequent resonance, simply from the changing scaling coefficients due to the changing $j$ of the resonance (Equations~\ref{eq:scalings first}). We indicate such a locus of $(J_0,\dot\beta)$ on Figure~\ref{fig:prob} as green crosses, for a particle with $e=0.01, \dot{a}=1$\,AU\,Myr$^{-1}$ for particles encountering external resonances with an Earth mass planet orbiting a Solar mass star at 1\,AU. Here the particle is migrating too fast to be captured into the 2:1 resonance, but is very likely to be captured into the 3:2, 4:3 and subsequent resonances. Capture probability decreases for higher $j$.

Our treatment does not include any eccentricity belonging to the massive planet. Correctly accounting for this requires including in the Hamiltonian another resonant term for the resonance involving the planet's longitude of pericentre (Q06). However, \cite{2006MNRAS.373.1245Q} showed that the strength of this resonant term can be reduced if the particle's osculating Keplerian eccentricity is replaced by the free secular eccentricity, which is approximately the same as the planet's eccentricity for particles close to the planet. Hence, for a first approximation, in the subsequent discussion we consider the impact of the planet's eccentricity as though it were the planetesimals'.

\subsection{Formation of resonances during planet formation}
\label{S:planets}

Here we discuss the capture of planets into resonances due to migration through a gas disc. Due to the complexity of the problem, this Hamiltonian model can only give us a crude understanding of this scenario. We have neglected the mass of one planet, so when the planets are of comparable masses it may be necessary to integrate the equations of motion for two planets (Equations~\ref{eq:H reduced parameter} and~\ref{eq:H reduced parameter 2nd order}). In spite of this limitation, the Hamiltonian model may give us some basic insight into the capture process.

There are several scenarios which may arise in a protoplanetary disc. For example, a giant planet undergoing type~II migration may capture planetesimals and low-mass embryos interior to it (e.g., \citealt{2007A&A...472.1003F}). The outer planets of our own Solar system may have once been in a chain of resonances, as \cite{2007AJ....134.1790M} suggested for initial conditions for the Nice model of the outer Solar System, and which \cite{2008A&A...482..333P} suggest may be a general outcome of migration of Jupiter and Saturn mass planets. Lower-mass super-Earths migrating inwards can encounter the exterior resonances of a giant planet (e.g., \citealt{2009MNRAS.397.1995P}). On the other hand, there may be a tendency for multi-planet systems to have the more massive planet on an external orbit \citep{2007Icar..191..158M}. Here we briefly discuss this latter case, where an inwardly migrating planet is capturing smaller planets or embryos in interior resonances.

On Figure~\ref{fig:adotcrit} (bottom panel) we have shown the typical migration rates due to Type~I migration (which is mass-dependent) and Type~II migration (mass-independent) for typical disc properties (MMSN, disc viscosity parametrised by $\alpha=0.01$, scale height $h=0.025$) at 1\,AU \citep{2009AREPS..37..321C}\footnote{We restrict ourselves to classical Type~I migration; the model can be used for the reduced or reversed migration found for example in non-isothermal discs \citep{2009A&A...506..971K} or isothermal discs in the non-linear regime \citep{2009MNRAS.394.2283P}, so long as a numerical or analytical migration rate is known.}. We see that, for the 2:1 resonance, Type~I migration at this location in this disc is always too rapid allow capture. Capture into the 2:1 resonance could occur however if the planets were located closer to the star at the epoch of capture. Although the Type~I migration rate decreases with planet mass, the critical migration rate decreases more strongly, so capture into this resonance is not possible with these disc parameters. Only more massive planets moving slowly under Type~II migration can capture into the 2:1 resonance, which will be an unavoidable outcome of migration in this disc if eccentricities are low, unless planets formed closer to each other than the 2:1 resonance. Lower mass planets can only capture into higher $j$ resonances. Indeed, we should see lower mass planets in higher $j$ resonances than higher mass planets if the Type I migration rate is correct. We also see that planets undergoing Type~I migration will be migrating too fast to capture smaller bodies into any second-order resonances, while planets undergoing Type~II migration will be able to capture smaller bodies into these resonances.

Next, we remark that the present libration amplitude of resonant bodies might have been imprinted at the instant of capture. First we consider how to achieve high libration amplitudes. For low eccentricity, both the Hamiltonian model and full numerical simulations \citep{2009A&A...497..595R} predict small libration amplitudes. \cite{2008ApJ...683.1117A} argue that turbulence in a disc, as well as breaking resonances completely, can lead to resonances with large libration amplitudes of $\gtrsim 60^\circ$, as seen in HD~128311, HD~82943, and HD~73526. Similar results were found by \cite{2009A&A...497..595R}. Here, we have shown that large libration amplitudes can also arise from high eccentricity at the instant of capture. For Jupiter-mass planets and first-order resonances the critical eccentricity to achieve such libration amplitudes is of order 0.1 (see figure~\ref{fig:amp} and equation~\ref{eq:scalings first}), and it is currently unclear whether a disc could pump planetary eccentricity to this level (see e.g., \citealt{2008Icar..193..475M}, \citealt{2010A&A...523A..30B}). However, for lower mass planets the necessary eccentricity is lower, and it may not be necessary to invoke turbulent fluctuations to explain a high libration amplitude if such low mass resonant planets are found to have high amplitudes of libration.

We can also consider how to achieve very low libration amplitudes. GJ~876 hosts two planets in a 2:1 resonance with both resonant arguments having very low libration amplitudes of $\sim 5^\circ$ \citep{2002ApJ...567..596L}. While \cite{2002ApJ...567..596L} obtained limits on planetary eccentricity at the time of capture of $e_\mathrm{b}\lesssim 0.06$, $e_\mathrm{c}\lesssim 0.03$ for planets b and c by requiring that capture into high order resonances not occur, we suggest that it may be possible to improve this: if the libration amplitude were very low at the time of capture, our results show that the eccentricities would have to be $e_\mathrm{b}\lesssim 0.01$, $e_\mathrm{c}\lesssim 0.02$.

These results ignore any subsequent evolution of the libration angle following capture. Libration amplitude decreases slightly with continued migration, but only very weakly ($\Delta\theta\propto t^{-1/8}$; see equation~\ref{eq:time evolution first}), so eccentricities at the moment of capture could have been slightly higher, depending on the extent of the migration. We also note that any eccentricity damping mechanism may reduce libration amplitudes, and we have already mentioned that stochastic fluctuations can increase it. 

There may be some caveats associated with using the simplified Hamiltonian model to model capture in a hydrodynamic disc. However, the full hydrodynamical simulations and those using an N-body integration with a prescribed migration rate give comparable results \citep[e.g.,][]{2004A&A...414..735K}. Since we have demonstrated the accuracy of our model against N-body integrations, we should not expect many significant differences between the Hamiltonian model and full hydrodynamic models. We also note that a large planet can open a gap in a disc and hence prevent other bodies from migrating into resonances that lie in the gap \citep{2008A&A...482..333P,2009MNRAS.397.1995P}, which will prevent capture into those resonances. {The capture process may also be affected by damping mechanisms such as gas drag or collisional damping, although these may be more important before and after resonance passage (for example, driving bodies towards the low-eccentricity regime prior to capture) than during capture itself.

Finally, we point out that our results may be useful for planetary population synthesis models which require a simple alternative to N-body integrations for planet--planet interactions \citep[e.g.,][]{2010ApJ...719..810I}.

\subsection{Debris disc structure}

\subsubsection{Planetesimal discs}

The non-axisymmetric structure seen in several debris discs has been ascribed to the migration of a planet into a planetesimal disc (e.g., W03, \citealt{2008A&A...480..551R}). We showed in the previous Section that the Hamiltonian model reproduces well the results of W03 who investigated this precise case.

Resonances affect the observed disc morphology because particles with nonzero eccentricity trapped in a resonance spend most of their time at specific longitudes relative to the planet \citep[W03;][]{2008A&A...480..551R}. This gives the appearance of clumps at these locations when the disc is imaged, with different resonances giving different numbers of clumps. They are seen so long as the libration amplitude is low enough that the clumps are not smeared out. If libration amplitudes are too high, the resonant dust particles appear to form a ring, axisymmetric save for a gap where the planet lies.

The Hamiltonian model is particularly useful for studying the resonant signatures of planets in debris discs because the fate of tens of thousands of particles encountering any first or second order resonance can be quickly determined, regardless of migration rates or particle eccentricities. Hence, many images of discs resulting from the migration of different sized planets into discs in different states of dynamical excitation can be easily made. Thus it will be particularly useful for exploring the disc structures generated by a wide range of possible perturbers for comparison to images expected in the near future from such projects as ALMA.

We now return to the case of a planet at around 40\,AU orbiting a star of 2.5 Solar masses. This is of interest because these parameters pertain to a hypothetical planet that may be imposing structure on the debris disc around Vega (W03). We shall discuss whether the planet causes a detectable signature in the disc if it migrates at 0.5\,AU\,Myr$^{-1}$, for varying planetary mass and eccentricity (as discussed above, we treat the planet's eccentricity as though it were the particles'). Figure~\ref{fig:pldisk} shows the effects of varying planet mass and eccentricity on the probability of capture and libration amplitude for 2:1 resonance, which have direct relevance for disc morphology. We see that libration amplitude increases significantly with eccentricity, as expected from figure~\ref{fig:amp}. The libration amplitude does not, however, depend strongly on planet mass. The capture probability, on the other hand, does, with the $1$ and $10\mathrm{M}_\oplus$ planet capturing very few particles at any eccentricity at this migration rate, although capture would occur if migration were slower. Thus, we can predict that a $1$ or $10\mathrm{M}_\oplus$ planet migrating at 0.5\,AU\,Myr$^{-1}$ will not create a detectable signature in a disc as its resonances fail to capture particles. Higher mass planets, however, will create signatures, with the observed structure depending on the eccentricity. As the eccentricity increases, capture probability goes down (Figure~\ref{fig:prob}) and libration amplitude increases (figure~\ref{fig:amp}). Thus, at low eccentricities, the planet will capture many particles with low libration amplitudes, which will cause well-defined clumps to be visible in the disc. At higher eccentricities, the number of particles captured will decrease and their libration amplitudes increase, leading to progressively weaker resonant signatures at higher eccentricities. This explains the results of \cite{2008A&A...480..551R}, who investigated this scenario with N-body integrations (see their figure~10). We also note that \cite{2008A&A...480..551R} found that there was little difference in observed structure regardless of whether the planet or the disc particles were eccentric, further justifying us modelling the planet's eccentricity as belonging to the particles.

\begin{figure}
\hspace{-.04\textwidth}
\includegraphics[width=.5\textwidth]{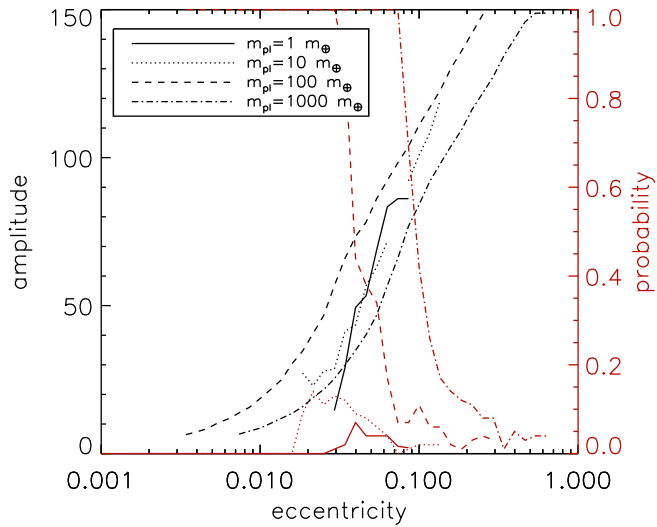}
\caption{Libration amplitudes in degrees (black lines) and capture probabilities (red lines) for planetesimals encountering the 2:1 resonance of a planet, orbiting a 2.5 Solar mass star, migrating outwards at $0.5$\,AU\,Myr$^{-1}$. Planet masses are shown in the legend.}
\label{fig:pldisk}
\end{figure}

It is worth noting that converting between the observed dust morphology and the underlying planetesimal population is not a trivial problem, since the resonances can affect the collisional evolution of planetesimals \citep{2006ApJ...639.1153W,2007CeMDA..99..169Q,2009ApJ...707..543S}, while dust grains can be liberated from a resonance by having a higher radiation pressure coefficient than their parents \citep{2006ApJ...639.1153W} or acquiring significant velocities relative to their parents as a result of the collisions which generated them \citep{2007A&A...462..199K}. A fully consistent disc model will need to take these effects into account. Since the dynamical behaviour of particles in the Hamiltonian model can be converted into physical positions and velocities, the Hamiltonian model can be used to generate a collisionless seed distribution for a collisional grooming algorithm to couple dynamical and collisional evolution \citep{2009ApJ...707..543S}. Currently such models use collisionless N-body integrations to generate the seed distribution. In the case of particles experiencing Poynting--Robertson drag, the Hamiltonian model will enable seed distributions for many particle sizes to be quickly generated, allowing a finer sampling of the dust size distribution.

These results may be applied in the Solar System to consider the implications of the observed libration widths of resonant KBOs for their eccentricities at the time of Neptune's migration. Figure~\ref{fig:amp} and Equations~\ref{eq:amp-t 1st} and~\ref{eq:amp-t 2nd} can be used to assess individual objects. However, a general point we can make already is that the observed high libration widths \citep{2007Icar..189..213L} would imply a high level of excitation, as already suggested by \cite{2003AJ....126..430C}.

These results can also be applied to consider the question of whether a disc can be stirred by a planet migrating towards the disc but too rapidly to capture bodies, which is known to be possible for diverging orbits \citep[e.g.,][]{2002ApJ...564L.105C}.

\subsubsection{Dust discs}
\label{S:zodi}

For less massive debris discs, such as the Solar System's zodiacal cloud and anticipated extrasolar analogues, particles can migrate substantial distances under PR drag before being destroyed by collisions. Note that currently observed debris discs are collision dominated and so the necessary migration cannot come from PR drag \citep{2005A&A...433.1007W}. Here we consider the scenario where the migration is due to PR drag acting on the dust, while the planet remains fixed.

As discussed previously, dust drifting under PR drag will encounter a succession of resonances. This is illustrated in Figure~\ref{fig:adotcrit}. The right hand axis of ordinates shows the dust particle size corresponding to a particular migration rate, given by $D=1.4L_\star/(\rho\dot{a})$, with the stellar luminosity $L_\star$ measured in Solar luminosities and the dust density $\rho$ measured in kg\,m$^{-3}$ \citep{1999ApJ...527..918W}\footnote{Note that because the PR migration rate depends on the particle's semi-major axis, a given migration rate corresponds to a slightly different particle size at different resonances. For this reason, the right-hand axis shows not particle size but size/$\alpha$.}. For $\rho=2500$\,kg\,m$^{-3}$, Figure~\ref{fig:adotcrit} predicts that a $1\mathrm{\,M}_\oplus$ planet at 1\,AU will be unable to capture low eccentricity particles of size less than about 10 micron into any resonances, but would capture particles of size 10--100 microns into high $j$ resonances, and particles of size $\gtrsim 100$ microns into only the 2:1 or 3:2 resonances.

In Figure~\ref{fig:zodi} we show the capture probabilities for 12 micron and 120 micron dust grains drifting under PR drag into the Earth's and Mars' resonances. Eccentricities were drawn from a uniform distribution between 0 and 0.2; the particles originate in the asteroid belt where the mean eccentricity is 0.15 \citep{1999ssd..book.....M} but will have had their eccentricities excited by crossing the $\nu_6$ secular resonance and damped by PR drag as they migrate. Particle sizes were drawn uniformly in the ranges $[11,13]$ microns and $[110,130]$ microns; note that the size distribution of the zodiacal dust at 1\,AU peaks at around 100 microns \citep{1993Sci...262..550L}. As expected, more of the larger particles are trapped than of the smaller ones. For particles of around 12 microns, with this eccentricity distribution, the Hamiltonian model predicts that the most populated resonances with Earth are the 5:4 and higher $j$, with lower $j$ resonances being less populated. This agrees qualitatively with the numerical study of \cite{1994Natur.369..719D}, who investigated trapping of 12 micron grains originating in the asteroid belt into Earth's resonances with an N-body code. They found that very few ($\lesssim 1\% $) particles are captured into the 4:3 and 5:4 resonances with Earth, rather more (2--3\%) into the 6:5 to 10:9 resonances, after which capture probaility decreases again (see Fig~4a of \citealt{1994Natur.369..719D}). Our capture probabilities are somewhat higher, which may be due to different eccentricity distributions or our neglect of stellar wind drag, but the picture of preferential trapping into $j\approx 6$ and higher is the same.

Mars, being less massive than the Earth, traps fewer particles than the Earth. Typically we find it only captures around 10--40\% of those Earth captures, depending on particle size and resonance $j$. This may explain why Mars lacks an observed resonant dust cloud while the Earth does not: too few particles are captured to yield a detectable signature. \cite{2000Icar..145...44K} searched without success for a cloud trailing Mars, finding an upper limit for its fractional overdensity of 18\% of Earth's cloud's overdensity, broadly consistent with the relative numbers of particles we find trapped into resonance with the two planets. If we had given particles encountering Earth's resonances lower eccentricity, which may be expected as they have had longer to be damped by PR drag, Mars would capture even fewer particles relative to the Earth. Martian capture probabilities may be further reduced due to the influence of the corotation resonance (W06), since the planet's eccentricity is the same order as that of the particles. A full investigation of this is beyond the scope of this paper, but it seems that Mars should have a resonant ring structure similar to that of Earth, at a level that may be close to the limits placed by current observations.

\begin{figure}
\includegraphics[width=.5\textwidth]{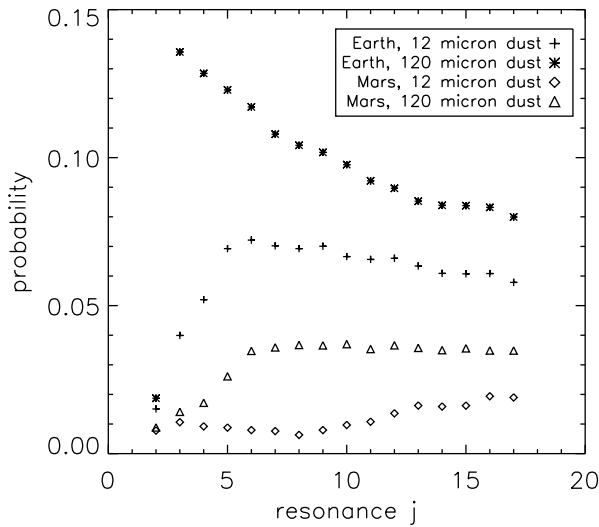}
\caption{Capture probabilities for 12 and 120 micron dust grains migrating into the Earth's and Mars' resonances, assuming particle eccentricities are randomly distributed in the range $[0,0.2]$.}
\label{fig:zodi}
\end{figure}

We intend to conduct a more thorough investigation of debris disc structure using our Hamiltonian model in a subsequent paper.

\subsection{Limitations}

We have included only one resonant term in our integrations. \cite{1996AJ....112.1278H} showed that resonances can be unstable when the additional resonant term associated with an eccentric planet is important, and Q06 showed that the probability of being captured can be reduced by the additional resonant term. However, since \cite{2008A&A...480..551R} found similar results in full N-body integrations when using eccentric particles and eccentric planets, including the extra resonant term explicitly may not be necessary.

We have restricted our numerical integrations to the case of a single planet plus a test particle. This may not adequately model two-planet systems where the planets are of comparable masses. In this case the equations given in the Appendix, \ref{eq:H reduced parameter} and \ref{eq:H reduced parameter 2nd order}, can be used. However, the increased number of free parameters will make a numerical study with a comparable grid resolution less practical.

We have neglected any eccentricity damping, which can occur together with change of semi-major axis when migrating in a gas disc or under PR drag, or experiencing dynamical friction in a planetesimal disc. This will limit the eccentricity growth due to continued migration in resonance. The effect on libration behaviour is dependent on the precise nature of the damping \citep{1995Icar..115...47G}.

We have only taken the lowest order term in the disturbing function, which is justified for low eccentricities. However, when eccentricities are high during migration, or are driven higher as migration continues after capture, higher order terms may become important. This is the case for the resonance between GJ\,876\,b and c \citep{2002ApJ...567..596L} and for the two libration centres of the 2:1 resonance, seen for eccentricities above $\sim 0.03$ \citep[W03;][]{2005ApJ...619..623M}. While post-capture eccentricity increase will not significantly affect the capture probability (a particle must cross a similar separatrix to enter the libration region in the first place) it will change the behaviour of libration amplitude and centre following capture.

We have not considered higher order resonances such as the 4:1, which can be important for very massive planets \citep{2003ApJ...588.1110K}.

\section{Summary and Conclusions}

We have systematically investigated the Hamiltonian model for capture of a test particle into first and second order resonances with a planet. The model reduces the full complexity of the restricted three body problem to only one degree of freedom, the resonant angle $\theta$, and two parameters, proportional to the particle's eccentricity and the migration rate. Only the relative migration rate of the two bodies, in the form $\dot{a}_1/a_1-\dot{a}_2/a_2$, is important. External and internal resonances behave the same way, but with different proportionality constants.

We confirmed previous work showing that capture into resonance is certain at low (rescaled) eccentricities and migration rates, possible with decreasing probability at high eccentricities, and impossible at low eccentricities and high migration rates. As eccentricity increases, the transition from certain capture at low migration rate to impossible capture at high migration rate broadens. At higher eccentricities, capture is possible with faster migration than at lower eccentricities. This effect is more pronounced for second order resonances.

We have also found the libration amplitudes and centres of the resulting resonant motion. The libration centres are offset from $\theta=\pi$, the centre in the absence of migration, by an amount increasing with the migration rate, agreeing with previous work. In addition, we have found that the offset is almost independent of eccentricity for first order resonances, except at high eccentricities. Libration amplitudes are small if capture occurs at low eccentricity and low migration rate. They are somewhat larger for migration rates just less than critical. Very large libration amplitudes ($\gtrsim 90^\circ$) can be attained if the initial eccentricity was very high.

We have also found the jumps in eccentricity when a particle encounters a resonance but is not captured. In the case of slow migration, the eccentricity is always driven down, in agreement with adiabatic theory. When the migration is fast and the eccentricity low, eccentricity jumps up. However, when the migration is fast and eccentricity high, eccentricity can jump either up or down.

We checked the results of the model against the N-body simulations of \cite{2003ApJ...598.1321W}, finding excellent quantitative agreement for capture probabilities, and qualitative agreement for libration amplitudes and offsets. We found that accounting for the particles' eccentricity is necessary to fully explain the dependence of capture probability on migration rate. While the effect of eccentricity has long been understood in the context of the Solar System, its implications for extrasolar planets are less well explored, and for low mass planets in particular its effects are important.

We have applied our model to several situations in which planet or particle migration is likely to occur. We can easily determine whether capture occurs for planets migrating in Type~I or Type~II regimes. The pre-capture eccentricity can be constrained by the present libration amplitude, with higher eccentricities giving higher libration amplitudes. For planets migrating through a gas disc, a non-zero eccentricity prior to capture can lead to large libration amplitudes such as those seen in the HD~128311 system. We find that, if planetary eccentricity can be raised to $e\gtrsim 0.1$ for Jupiter-mass planets, for example by planet-disc interaction, the resonant capture process by itself will result in high libration amplitudes without the need to invoke extra mechanisms such as turbulent torques. For lower mass planets the necessary eccentricity is lower. Also, very low libration amplitudes (e.g., GJ~876) suggest a low eccentricity during the capture process. We also predict, based on classical formulae for migration rates, that lower mass planets will be found in higher $j$ first order resonances than higher mass planets, and planets migrating in the type~I regime will be moving too fast to capture smaller particles into any second-order resonance. The model may also be useful for population synthesis of multi-planet systems where it is necessary to account for planet--planet interactions without recourse to full N-body integrations.

We then discussed debris disc structure. A planet that has migrated into a disc will impose clumpy structure on a dynamically cold disc, but the clumps will be at a lower level and smeared out for migration into excited discs. For dust migrating under PR drag, the model can explain the structure of the Earth's resonant ring reasonably well. We predict that Mars has a dust ring at a level $\lesssim 25\%$ that of Earth's, consistent with observed upper limits.

The data on capture probabilities, libration amplitudes and offsets, and eccentricity jumps, are available on-line at \url{http://www.ast.cam.ac.uk/~ajm233/} and  at the journal website and may be used provided that this work is cited.

\section*{Acknowledgements}

AJM is grateful for the support of an STFC studentship. The authors would like to thank Alice Quillen for useful discussions and comments on the paper, the referee John Chambers for useful suggestions, and Nikhil Shah for early work on this project.

\appendix

\section{Hamiltonian Model}

Here we detail the mathematical derivation of the scalings from physical variables to dimensionless variables, and some properties of the Hamiltonian model such as the evolution of libration amplitude and offset with time. Although many of these results may be found in the literature, we think it useful to gather them together for reference.

First we derive the scalings to link physical variables such as eccentricity to the dimensionless ones such as the momentum $J$. This closely follows Q06; we generalise to consider two massive planets.

We use Poincar\'e's canonical variables: the generalised momenta are
\begin{eqnarray}
\Lambda_i & = & m_i\sqrt{\mu_ia_i} \nonumber\\
\Gamma_i & = & \Lambda_i\left(1-\sqrt{1-e_i^2}\right),\label{eq:Poincare}
\end{eqnarray}
which are conjugate to the mean longitudes $\lambda_i$ and the negative of the longitudes of pericentre $\varpi_i$ respectively. Here, $\mu_i=\mathcal{G}(m_\star+m_i)$. The Keplerian part of the Hamiltonian is therefore
\begin{equation}\label{eq:H_kep}
\mathcal{H}_\mathrm{kep} = -\frac{m_1^3\mu_1^2}{2\Lambda_1^2} -\frac{m_2^3\mu_2^2}{2\Lambda_2^2}.
\end{equation}
We work with low eccentricities, so that $\Gamma_i/\Lambda_i\sim e_i^2/2$.

We now non-dimensionalise as follows. Distances are put in units of $a_0$, the semi-major axis of the inner planet when the nominal resonance is reached. Times are put in units of the inverse of the mean motion of the inner planet at that point. Masses are put in units of the mass of the inner planet. With this choice of units, the Keplerian Hamiltonian is
\begin{equation}\label{eq:H_kep non-dim}
\mathcal{H}_\mathrm{kep}^\prime = -\frac{1}{2\Lambda_1^{\prime2}} -\frac{m_2^{\prime3}\mu_2^{\prime2}}{2\Lambda_2^{\prime2}},
\end{equation}
with primes denoting the non-dimensional quantities. We now drop the primes, and make the approximation $\mu_2\approx\mu_1=1$, so that
\begin{equation}\label{eq:H_kep non-dim simple}
\mathcal{H}_\mathrm{kep} = -\frac{1}{2\Lambda_1^2} -\frac{\nu^3}{2\Lambda_2^2},
\end{equation}
where $\nu=m_2/m_1$ is the ratio of the two planets' masses.

To study the dynamics as the system passes through resonance, we expand about the nominal resonance location where $\Lambda_i=\Lambda_{i,0}$, with $\Lambda_{1,0}=1$ and $\Lambda_{2,0}=\nu\alpha_0^{-1/2}$, $\alpha_0=[(j-k)/j]^{2/3}$ being the semi-major axis ratio at the nominal resonance. This gives the Keplerian Hamiltonian
\begin{eqnarray}
\mathcal{H}_\mathrm{kep} &=& I_1 - \frac{3}{2}I_1^2 + \frac{j-k}{j}I_2 - \frac{3}{2}\left(\frac{j-k}{j}\right)^{4/3}\nu^{-1}I_2^2\nonumber\\
&=&I_1 - \frac{3}{2}I_1^2 + \alpha_0^{3/2}I_2 - \frac{3}{2}\alpha_0^2\nu^{-1}I_2^2,\label{eq:H_kep 2nd order}
\end{eqnarray}
where $I_i=\Lambda_i-\Lambda_{i,0}$ measures the distance from resonance. This expansion assumes $I_2\alpha^{1/2}/\nu\ll 1$. This is satisfied for all $\nu$ since $I_2 = \mathcal{O}(\nu)$ as $\nu\to 0$. We just require the bodies to be close to resonance in semi-major axis.

From the disturbing function we retain the relevant resonant terms. The resonant Hamiltonian is
\begin{eqnarray}
\mathcal{H}_\mathrm{res}&=&\sum_{p=0}^k r_{k,p}\Gamma_1^{p/2}\Gamma_2^{(k-p)/2}\nonumber\\
&\times&\cos\left[j\lambda_2-(j-k)\lambda_1-p\varpi_1-(k-p)\varpi_2\right].\label{eq:H_res}
\end{eqnarray}
Again, the $r$ coefficients are functions of $\alpha$ and regarded as constant. In terms of the $f_i$ functions, they are
\begin{eqnarray}
r_{1,0}&=&-\nu^{1/2}\mu\left(\sqrt{2}\alpha^{5/4}f_{31}(\alpha)-2\sqrt{2}\alpha^{9/4}\mathbb{I}(2:1)\right)\nonumber\\
r_{1,1}&=&-\sqrt{2}\nu\mu\alpha f_{27}(\alpha),\label{eq:r coef}
\end{eqnarray}
where $\mathbb{I}(2:1)=1$ for the 2:1 resonance and is zero otherwise; this is from the indirect disturbing function.

The $f_i$ functions are tabulated for first-order resonances in Table~(\ref{tab:f_i k=1}).

\begin{table}
  \begin{center}
    \begin{tabular}{cccccc}
      $j$ & $\alpha$  & $f_{27}$  & $f_{31}$\\\hline\\
      2   & 0.629961  & -1.19050 & 1.68831\\
      3   & 0.763143  & -2.02523 & 2.48401\\
      4   & 0.825482  & -2.84043 & 3.28326\\
      5   & 0.861774  & -3.64962 & 4.08371\\
      6   & 0.885549  & -4.45614 & 4.88470\\
      7   & 0.902337  & -5.26125 & 5.68601\\
      8   & 0.914826  & -6.06553 & 6.48749\\
      9   & 0.924482  & -6.86925 & 7.28909\\
      10  & 0.932170  & -7.67260 & 8.09077\\
      11  & 0.938437  & -8.47572 & 8.89251
    \end{tabular}
    \caption{Values of $\alpha$ and $f_i$ for first-order resonances.}
    \label{tab:f_i k=1}
  \end{center}
\end{table}

\subsection{First order resonances}

Now we consider first-order resonances. There are two resonant angles $\theta_1$ and $\theta_2$; we effect a point transformation to these angles:
\begin{eqnarray}
\theta_1&=&j\lambda_2+(1-j)\lambda_1-\varpi_1\nonumber\\
\theta_2&=&j\lambda_2+(1-j)\lambda_1-\varpi_2\nonumber\\
\theta_3&=&\lambda_1\nonumber\\
\theta_4&=&\lambda_2.\label{eq:theta 1-4}
\end{eqnarray}
The new momenta satisfy
\begin{eqnarray}
(1-j)(J_1+J_2)+J_3&=&I_1\nonumber\\
j(J_1+J_2)+J_4&=&I_2\nonumber\\
J_1&=&\Gamma_1\nonumber\\
J_2&=&\Gamma_2\label{eq:J 1-4}.
\end{eqnarray}
The Keplerian Hamiltonian is then
\begin{eqnarray}
\mathcal{H}_\mathrm{kep}&=&-3(1-j)J_3(J_1+J_2)-\frac{3}{2}(1-j)^2(J_1+J_2)^2\nonumber\\
 &&-\frac{3j\alpha^2}{\nu}J_4(J_1+J_2)-\frac{3\alpha^2}{2\nu}j^2(J_1+J_2)^2\label{eq:H_kep J}.
\end{eqnarray}
The resonant Hamiltonian is
\begin{equation}\label{eq:H_res J}
\mathcal{H}_\mathrm{res}=r_{1,0}J_2^{1/2}\cos\theta_2 + r_{1,1}J_1^{1/2}\cos\theta_1.
\end{equation}
This reduces the system to two degrees of freedom: the mean longitudes $\theta_3$ and $\theta_4$ do not appear in the equations. The momenta $J_3$ and $J_4$ are therefore constants; we dropped terms involving only these momenta from Equation~(\ref{eq:H_kep J}). We simulate migration by explicitly varying the momenta $J_3$ and $J_4$ with time. We have
\begin{equation}\label{eq:J_3 dot}
\dot{J}_3\approx\dot{I}_1\approx\dot{a}_1/2.
\end{equation}
and
\begin{equation}\label{J_4 dot}
\dot{J}_4\approx\dot{I}_2\approx\nu\dot{a}_2\left(\frac{j-1}{j}\right)^{1/3}/2.
\end{equation}

\subsubsection{Reducing number of parameters}

We return to the general Hamiltonian (Equations~\ref{eq:H_kep J}--\ref{eq:H_res J}). This may be written
\begin{eqnarray}
\mathcal{H}&=&\alpha J_1 + \beta J_2 + \gamma J_1^2 + 2\gamma J_1J_2 + \gamma J_2^2 \nonumber\\
&&+s_3J_1^{1/2}J_2^{1/2}\cos(\theta_1-\theta_2) + r_{1,1}J_1^{1/2}\cos\theta_1 \nonumber\\
&&+ r_{1,0}J_2^{1/2}\cos\theta_2,\label{eq:H full parameter}
\end{eqnarray}
where
\begin{eqnarray}
\alpha&=&-3\left[(1-j)J_3+\frac{j\alpha_0^2}{\nu}J_4\right] + s_{21}\nonumber\\
\beta&=&-3\left[(1-j)J_3+\frac{j\alpha_0^2}{\nu}J_4\right] + s_{22}\nonumber\\
\gamma&=&-\frac{3}{2}\left[(1-j)^2+\frac{\alpha_0^2j^2}{\nu}\right]\label{eq:parameters 1}.
\end{eqnarray}
These six parameters we reduce to four by rescaling momentum:
\begin{eqnarray}
\mathcal{H}^\prime&=&\alpha^\prime J_1^\prime + \beta^\prime J_2^\prime + (J_1^\prime+J_2^\prime)^2\nonumber\\
&&+s^\prime J_1^{\prime 1/2}J_2^{\prime 1/2}\cos(\theta_1-\theta_2) - J_1^{\prime 1/2}\cos\theta_1 \nonumber\\
&&+r^\prime J_2^{\prime 1/2}\cos\theta_2,\label{eq:H reduced parameter}
\end{eqnarray}
where
\begin{eqnarray}
\mathcal{H}^\prime&=&-\left|\frac{\gamma^{1/3}}{r_{1,1}^{4/3}}\right|\mathcal{H}\nonumber\\
J_i^\prime&=&\left|\frac{\gamma}{r_{1,1}}\right|^{2/3}J_i\nonumber\\
\alpha^\prime&=&-\alpha r_{1,1}^{-2/3}\left|\gamma\right|^{-1/3}\nonumber\\
\beta^\prime&=&-\beta r_{1,1}^{-2/3}\left|\gamma\right|^{-1/3}\nonumber\\
r^\prime&=&-r_{1,0}/r_{1,1}.
\end{eqnarray}
and the new time
\begin{equation}\label{eq:tprime}
t^\prime = t\gamma^{1/3}r_{1,1}^{2/3}.
\end{equation}
Note that $\gamma<0$, and $r_{1,1}>0$ for first-order resonances.

We may write the parameters in terms of physical quantities $\nu,\mu,\alpha_0$:
\begin{eqnarray}
\alpha^\prime&=&\frac{3(j-1)\left(\sqrt{a_2/a_{2,0}}-\sqrt{a_1}\right)+2\nu\mu\alpha_0 f_2}
{3^{1/3}\left(\nu\mu\alpha_0 f_{27}\right)^{2/3}\left[(1-j)^2+\frac{\alpha_0^2j^2}{\nu}\right]^{1/3}}\nonumber\\
\beta^\prime&=&\frac{3(j-1)\left(\sqrt{a_2/a_{2,0}}-\sqrt{a_1}\right)+2\mu\alpha_0^{3/2} f_2}
{3^{1/3}\left(\nu\mu\alpha_0 f_{27}\right)^{2/3}\left[(1-j)^2+\frac{\alpha_0^2j^2}{\nu}\right]^{1/3}}\nonumber\\
r^\prime&=&-\frac{\alpha_0^{1/4}f_{31}-2\alpha_0^{5/4}\mathbb{I}(2:1)}{\nu^{1/2}f_{27}}.\\
t^\prime&=&3^{1/3}\left(\nu\mu\alpha_0 f_{27}\right)^{2/3}\left[(1-j)^2+\frac{\alpha_0^2j^2}{\nu}\right]^{1/3}t
\end{eqnarray}

Migration entails varying the coefficients $\alpha^\prime$ and $\beta^\prime$ with time. We have:
\begin{equation}\label{eq:alphaprimedot}
\frac{\mathrm{d}{\alpha}^\prime}{\mathrm{d}t^\prime}=\frac{\mathrm{d}{\beta}^\prime}{\mathrm{d}t^\prime}
=\frac{3^{1/3}\left(\frac{\dot{a}_2}{a_2}-\frac{\dot{a}_1}{a_1}\right)}{2(j-1)^{1/3}\left(\frac{m_2}{m_0}\alpha_0 f_{27}\right)^{4/3}\left(1+\frac{m_1}{\alpha_0 m_2}\right)^{2/3}}.
\end{equation}

\subsubsection{Limiting case 1: outer test particle and zero eccentricity planet}

We consider the limit $\nu\to 0$. This is relevant for either a planet capturing planetesimals as it migrates outwards, or dust spiralling inwards under P.\,R.\ drag. In this case a more convenient scaling is using $r_{1,0}$ rather than $r_{1,1}$. The non-constant terms in the Hamiltonian are
\begin{equation}\label{eq: H outer}
\mathcal{H}^\prime = J_2^{\prime 2} + \beta_0J_2^\prime - J_2^{\prime 1/2} \cos\theta_2,
\end{equation}
where
\begin{equation}\label{eq:beta0}
\beta_0=\frac{3(j-1)\left(\sqrt{a_2/a_{2,0}}-\sqrt{a_1}\right)+2\mu\alpha_0^{3/2} f_2}
{\left\{3\mu^2\alpha_0^{9/2}j^2\left[f_{31}-2\alpha_0\mathbb{I}(2:1)\right]^2\right\}^{1/3}}.
\end{equation}
Note that all terms in Equation~(\ref{eq: H outer}) are $\mathcal{O}(\nu^0)$. For a migrating planet, the migration rate is given by
\begin{equation}\label{eq:beta0dot}
\frac{\mathrm{d}\beta_0}{\mathrm{d}t^\prime} = \frac{3^{1/3}(j-1)\left(\dot{a}_2/a_{2,0}-\dot{a}_1\right)}
{2\alpha_0^3\left\{ \mu j \left[f_{31}-2\alpha_0\mathbb{I}(2:1)\right] \right\}^{4/3}}.
\end{equation}

\subsubsection{Limiting case 2: inner test particle and zero eccentricity planet}

The limit $\nu\to\infty$, $\mu\to 0$, $\mu\nu=m_2/m_0$ constant, is relevant for a planet capturing planetesimals as it migrates inwards. The Hamiltonian is
\begin{equation}\label{eq:H inner}
\mathcal{H}^\prime = J_1^{\prime 2} + \alpha_\infty J_1^\prime - J_1^{\prime 1/2}\cos \theta_1
\end{equation}
where
\begin{equation} \label{eq:alpha infinity}
\alpha_\infty=\frac{3(j-1)\left(\sqrt{a_2/a_{2,0}}-\sqrt{a_1}\right)+2\mu\nu\alpha_0 f_2}
{3^{1/3}\left[\mu\nu\alpha_0 f_{27}(1-j)\right]^{2/3}}.
\end{equation}
All terms in Equation~(\ref{eq:H inner}) are $\mathcal{O}(\nu^0)$. The migration rate in this case is given by
\begin{equation}\label{eq:alphainfdot}
\frac{\mathrm{d}\alpha_\infty}{\mathrm{d}t^\prime} = \frac{3^{1/3}(\dot{a}_2/a_{2,0}-\dot{a}_1)}{2(j-1)^{1/3}\left(\nu\mu \alpha_0 f_{27}\right)^{4/3}}.
\end{equation}

\subsection{Second order resonances}

For second order resonances there are three resonant terms. The scaled Hamiltonian (\emph{cf.} Equation \ref{eq:H reduced parameter}) is
\begin{eqnarray}
\mathcal{H}^\prime&=&\alpha^\prime J_1^\prime + \beta^\prime J_2^\prime + (J_1^\prime+J_2^\prime)^2\nonumber\\
&&+s^\prime J_1^{\prime 1/2}J_2^{\prime 1/2}\cos(\theta_1-\theta_2) - J_1\cos 2\theta_1 \nonumber\\
&&+r_1^\prime J_1^{\prime 1/2}J_2^{\prime 1/2}\cos(\theta_1-\theta_2)+r_2^\prime J_2\cos 2\theta_2,\label{eq:H reduced parameter 2nd order}
\end{eqnarray}
with the rescaled time
\begin{equation}\label{eq: tprime second order}
t^\prime = 2\mu\nu\alpha_0f_{45}t,
\end{equation}
the rescaled momenta
\begin{equation}\label{eq: jprime second order}
J_i^\prime = \frac{3\left[(2-j)^2+\frac{\alpha_0^2j^2}{\nu}\right]}{16\mu\nu\alpha_0f_{45}}\Gamma_i,
\end{equation}
and the parameters
\begin{eqnarray}
\alpha^\prime & = & \frac{3\left[(2-j)a_1^{1/2}+j\alpha_0^2a_2^{1/2}\right] + 4\mu\nu\alpha_0f_2}{4\mu\nu\alpha_0f_{45}}\label{eq:alphaprime second order}\\
\beta^\prime & = & \frac{3\left[(2-j)a_1^{1/2}+j\alpha_0^2a_2^{1/2}\right] + 4\mu\alpha_0^{3/2}f_2}{4\mu\nu\alpha_0f_{45}}\label{eq:betaprime second order}\\
r_1^\prime & = & -\frac{\alpha_0^{1/4}f_{49}}{\nu^{1/2}f_{45}}\label{eq:r1prime second order}\\
r_2^\prime & = & -\frac{\alpha_0^{1/2}\left[2f_{53}-\frac{27}{4}\alpha_0\mathbb{I}(3:1)\right]}{2\nu f_{45}}.\label{eq:r2prime second order}
\end{eqnarray}
The migration rate is given by
\begin{equation}\label{eq:dbdt second order}
\frac{\mathrm{d}\alpha^\prime}{\mathrm{d}t^\prime}=\frac{\mathrm{d}\beta^\prime}{\mathrm{d}t^\prime}=-\frac{3(j-2)(\dot{a}_2/a_2-\dot{a}_1)}{16\mu^2\nu^2\alpha_0^2f_{45}^2}.
\end{equation}

The values of the $f_i$ for second-order resonances are tabulated in Table~(\ref{tab:f_i k=2}).

\begin{table}
  \begin{center}
    \begin{tabular}{cccccc}
          $j$ & $\alpha$ & $f_{45}$  & $f_{53}$ \\\hline\\
          3   & 0.480750 & 0.598759 & 1.98591 \\
          5   & 0.711379 & 3.27381  & 5.68728 \\
          7   & 0.799064 & 7.87052  & 11.3173 \\
          9   & 0.845740 & 14.3866  & 18.8674 \\
          11  & 0.874782 & 22.8216  & 28.3365 \\
          13  & 0.894608 & 33.1755  & 39.7246 \\
          15  & 0.909009 & 45.4473  & 53.0314 \\
          17  & 0.919944 & 59.6382  & 68.2557 \\
          19  & 0.928532 & 75.7482  & 85.3994 \\
          21  & 0.935455 & 93.7765  & 104.462
    \end{tabular}
    \caption{Values of $\alpha$ and $f_i$ for second-order resonances.}
    \label{tab:f_i k=2}
  \end{center}
\end{table}

For an outer test particle and a planet on a circular orbit we instead rescale by $r_{2,0}$, giving
\begin{eqnarray}
t^\prime&=&\alpha_0^{3/2}\mu\left[2f_{53}-\frac{27}{4}\alpha_0\mathbb{I}(3:1)\right]t \label{eq:tprime 2nd order ext}\\
J_2^\prime&=&\frac{3j^2}{16\mu\left[2f_{53}-\frac{27}{4}\alpha_0\mathbb{I}(3:1)\right]}e_2^2 \label{eq:j2prime 2nd order ext}\\
\beta^\prime&=&\frac{3\left[(2-j)a_1^{1/2}+j\alpha_0^2a_2^{1/2}\right] + 4\mu\alpha_0^{3/2}f_2}{2\alpha_0^{3/2}\mu\left[2f_{53}-\frac{27}{4}\alpha_0\mathbb{I}(3:1)\right]} \label{eq:betaprime 2nd order ext}\\
\frac{\mathrm{d}\beta^\prime}{\mathrm{d}t^\prime}&=&\frac{3(j-2)(\dot{a}_2/a_2-\dot{a}_1)}{4\alpha_0^3\mu^2\left[2f_{53}-\frac{27}{4}\alpha_0\mathbb{I}(3:1)\right]^2}. \label{eq:betadot 2nd order ext}
\end{eqnarray}

\subsection{Evolution in resonance}\label{s: evolution analytical}

\subsubsection{First-order resonances}

After a particle has been captured into resonance, the trajectory continues to evolve so long as the imposed migration is still present. An analytical description of the behaviour is possible when $|\beta|$ is sufficiently large. Both the amplitude and the offset of the libration centre of the libration in $\theta$ decrease with time, while the momentum $J$ increases.

For first order resonances, the libration centre can be found by considering the equations of motion for $\theta$ and $J$:
\begin{eqnarray}
\dot{\theta} &=& 2J+\beta-\frac{1}{2}J^{-1/2}\cos \theta \label{eq:thetadot 1st}\\
\dot{J} &=& J^{1/2}\sin \theta \label{eq:Jdot 1st},
\end{eqnarray}
whence,
\begin{eqnarray}
\ddot{\theta}&=&2J^{1/2}\sin\theta+\dot\beta+\frac{1}{4}J^{-1}\sin\theta\cos\theta\nonumber\\
&&+\frac{1}{2}J^{-1/2}\left(2J+\beta-\frac{1}{2}J^{-1/2}\cos\theta\right)\sin\theta.\label{eq:theta double dot 1st}
\end{eqnarray}
Now, the elliptical fixed point occurs at approximately $J_\mathrm{eq}=-\beta/2$ for large $|\beta|$, thus giving the increase in momentum. For large $|\beta|$ and small $\theta$, Equation~\ref{eq:theta double dot 1st} tells us that
\begin{equation}\label{eq:theta double dot 1st simple}
\ddot{\theta}=\sqrt{-2\beta}\theta+\dot{\beta},
\end{equation}
so that the libration centre offset is given by
\begin{equation}\label{eq:centre 1st}
\theta_\mathrm{eq}=-\frac{\dot{\beta}}{\sqrt{-2\beta}}.
\end{equation}

To find the evolution of the libration amplitude, we consider the equations of motion in Cartesian coordinates, following the canonical transformation
\begin{equation}\label{eq:poincare cartesian}
x=\sqrt{2J}\cos{\theta},\qquad y=\sqrt{2J}\sin{\theta}.
\end{equation}
We calculate the libration amplitude by imposing a constant area on the small elliptical trajectory enclosed by the orbit; a knowledge of the aspect ratio of the ellipse then suffices to determine the amplitude in $y$, and hence $\theta$.

We improve the estimate for the equilibrium point by a single iteration of the Newton-Raphson method,\footnote{Note that working directly with $J_\mathrm{eq}=-\beta/2$ leads to the linear terms in the equation for $\dot X$ vanishing.} giving
\begin{equation}\label{eq:x_eq 1st}
x_\mathrm{eq}\sim\alpha+1/(2\sqrt{2}\alpha^2),
\end{equation}
where $\alpha=\sqrt{-\beta}$.
Writing $X=x-x_\mathrm{eq}$ and linearising gives
\begin{equation}\label{eq:linear eqq 1st}
\dot X = -y/\sqrt{2}\alpha,\qquad
\dot y = 2\alpha^2 X
\end{equation}
to leading order in $\alpha$. This describes motion in an ellipse about $\theta=0$, with ratio of semi-axes $\Delta X/\Delta y=2^{-3/4}\alpha^{-3/2}$. The area enclosed by a trajectory is thus $A \propto \alpha^{-3/2} (\Delta y)^2$. Since $\Delta\theta\sim\Delta y/x$, and $x\sim\alpha$, we find
\begin{equation}\label{eq:libwidth 1st}
\Delta\theta\sim C \alpha^{-1/4} \propto \beta^{-1/8}
\end{equation}
for some constant $C$. This constant is not determined by this analysis; we obtain it numerically in \S\,\ref{S:numerics}.

This also gives the period of libration:
\begin{equation}\label{eq:period 1st}
t_\mathrm{lib}=2^{3/4}\pi|\beta|^{-1/4}.
\end{equation}

\subsubsection{Second-order resonances}

For second-order resonances, a similar analysis gives 
\begin{equation}\label{eq:centre 2nd}
\theta_\mathrm{eq}=\frac{\dot{\beta}}{4\beta},
\end{equation}
\begin{equation}\label{eq:amp 2nd}
\Delta\theta\propto|\beta|^{-1/4},
\end{equation}
and
\begin{equation}\label{eq:period 2nd}
t_\mathrm{lib}\approx\pi|\beta|^{-1/2}.
\end{equation}

\bibliographystyle{/home/ajm233/latex/mnras/mn2e}
\bibliography{bibliography}

\end{document}